\documentclass[12pt,preprint]{aastex}

\usepackage{natbib}
\newcommand{\totj}{\frac{t}{\tau_j}}
\newcommand{\emtotj}{e^{-t/\tau_j}}
\newcommand{\emtoti}{e^{-t/\tau_i}}
\newcommand{\muij}{\mu_{ij}}
\newcommand{\hinv}{h^{-1}}

\begin{document}


\title{Exploring the Energetics of Intracluster Gas with a Simple and
Accurate Model}


\author{Paul Bode and Jeremiah P. Ostriker} 
\affil{Department of Astrophysical Sciences, Peyton Hall, Princeton
University, Princeton, NJ 08544; 
bode@astro.princeton.edu, ostriker@princeton.edu}
\author{Alexey Vikhlinin}
\affil{Harvard-Smithsonian Center for Astrophysics, 60 Garden St., 
Cambridge, MA 02138; avikhlinin@cfa.harvard.edu}



\begin{abstract}
The state of the hot gas in clusters of galaxies is investigated with 
a set of model clusters, created by assuming a polytropic equation of
state ($\Gamma=1.2$)
and hydrostatic equilibrium inside gravitational potential wells
drawn from a dark matter simulation.  Star formation, energy input,
and nonthermal pressure support are included.  To match the gas
fractions seen in non-radiative hydrodynamical simulations, roughly
5\% of the binding energy of the dark matter must be transferred to
the gas during cluster formation;  the presence of nonthermal pressure
support increases this value.  In order to match X-ray observations,
scale-free behavior must be broken.  This can be due to either variation
of the efficiency of star formation with cluster mass $M_{500}$, or the
input of additional energy proportional to the formed stellar mass
$M_F$.  These two processes have similar effects on X-ray scalings.
If 9\% of the gas is converted into stars, independent of cluster mass,
then feedback energy input of $1.2\times 10^{-5}M_Fc^2$ 
(or $\sim 1.0$ keV per particle) is required to
match observed clusters.  Alternatively, if the stellar mass fraction varies as
$\propto M_{500}^{-0.26}$ then a lower feedback of 
$4\times 10^{-6}M_Fc^2$ is needed, and if the stellar fraction varies as steeply
as $\propto M_{500}^{-0.49}$ then
no additional feedback is necessary.  The model clusters reproduce the
observed trends of gas temperature and gas mass fraction with cluster
mass, as well as observed entropy and pressure profiles; thus they
provide a calibrated basis with which to interpret upcoming SZ surveys.
One consequence of the increased gas energy is that the baryon fraction
inside the virial radius is $\la 90$\% of the cosmic mean,
even for the most massive clusters.
\end{abstract}



\section{Introduction}   \label{sec:intro}

In the near future, clusters of galaxies will prove to be an 
increasingly important probe into the abundance and properties of both
dark matter and dark energy.
Data on large numbers of clusters are becoming available from
surveys at many wavelengths:
in the optical/infrared, from galaxy overdensities
\citep{Koesterea07,YeeGGMHE07,Eisenhardtea08,LopesCKJ09}
or from weak gravitational lensing of
background objects \citep{Dahle07z}; 
in the X-ray 
\citep{Bohringerea07,BureninVHEQM07,Finoguenovea07,Pacaudea07};
and in the microwave, by locating temperature decrements
caused by the  Sunyaev-Zeldovich (SZ) effect
\citep{Dobbsea06,Kosowsky06,MalteSchaferB07,Muchovejea07,LinWUKLNHLWH08,BonamenteJLCNM08,Staniszewskiea08z}.

Properly interpreting such surveys will require a sound
understanding of the state of the Intra-Cluster Medium (ICM).
Based on the assumption that the thermal energy in the gas
comes solely from gravitational collapse, one may derive
expected self-similar scalings between halo mass and temperature,
X-ray luminosity, etc.\citep{Kaiser91,BryanN98,EkeNF98}.
However, these scalings do not match the actual behavior of observed clusters 
\citep[e.g][]{ArnaudEvrard99,VikhlininKFJMMVS06,ArnaudPP07,PrattCAB08z,SunVDJFV09,VikhlininBEFHJKMNQV09}.
Additional modification of the thermal state of the gas from
non-gravitational sources will bring these predictions into
better agreement with observed scalings \citep{Kaiser91}.
One such modification is star formation; by removing gas with
the shortest cooling times, star formation leaves behind gas
with a higher mean entropy 
\citep{VoitBryan01,TozziNorman01,VoitBBB02}.
Energy injection from preheating or stellar and AGN feedback likewise
affects scalings, as numerous simulations which include such feedback
have shown
\citep[e.g.][and references therein]{EttoriDBM06,
BorganDMCSDMTTT06,MuanwongKT06,NagaiKV07,KaydSABLPST07,SijackiSMH07,
BhattacharyaMK08,BorganiDDS08,DaveOS08,PuchweinSS08,ShortT08z}.
The importance of these processes can be investigated
analytically by considering hydrostatic, and usually polytropic,
gas residing in various dark matter potentials
\citep{SutoSM98,BaloghBP99,WuFN00,Loewenstein00,
TozziNorman01,KomatsuSeljak01,BabulBLP02,VoitBBB02,
DosSantosDore02,AscasibarYMS03,ShimizuKSS04,LapiCM05,AfshordiLS05,
SolanesMGS05,AscasibarD08,BowerMB08,CiottiP08,MoodleyWGT08z}.

\defcitealias{OstrikerBB05}{Paper I}
\defcitealias{BodeOWS07}{Paper II}

\citet[][hereafter Paper I]{OstrikerBB05} and \citet[][Paper II]{BodeOWS07}
developed a simple method for populating the halo Dark Matter (DM)
potentials from N-body simulations with gas, based on the assumptions
of hydrostatic equilibrium and a polytropic equation of state.
This method can be calibrated against local X-ray observations,
and then be used to create realistic sky maps \citep{SehgalTHB07}.  
Rather than assuming a simple analytic profile, the distributions
of halo concentration, substructure,
and triaxiality are included, as well as
any spatial correlations, since the halos are drawn
from an accurate  large scale dark matter simulation.
This procedure has advantages over performing full
hydrodynamic simulations.
Because less computational expense is required,
large volumes can be simulated.  Also, one can vary parameters, such as
the amount of star formation and feedback, without having
to redo an entire simulation.

This paper presents a number of improvements
to the model presented in
\citetalias{OstrikerBB05} and \citetalias{BodeOWS07}.
A more sophisticated description of star
formation is included, which accounts for gas recycling
and allows for a calculation of metallicity evolution.
\citetalias{BodeOWS07} assumed that every cluster halo has
the same stellar mass fraction; here we allow this fraction to vary
with halo mass.
Also \citetalias{BodeOWS07} did not allow for transfers of
energy from the dark matter to the gas.  It is evident from
simulations that this does occur, so we allow for this transfer here.
Because of these limitations, \citetalias{BodeOWS07} required a
large amount of feedback energy to match X-ray observations.
As we shall see, the improvements presented here  significantly
reduce the feedback required.

With this method, we examine how star formation and different forms
of energy input affect the ICM.  It is possible to produce model
clusters which match existing observational constraints, and thus
make predictions for upcoming surveys.
The method is described in Sec.~\ref{sec:method} and
Appendix \ref{sec:starform}.  
In Sec.~\ref{sec:feedback} it is then
applied to a simulation reflecting the
current best-fit cosmological parameters and the results
are compared to X-ray data; in particular the effects of varying star
formation and feedback are examined.
In Sec.~\ref{sec:compother} a standard model is compared
to further X-ray and SZ observation; we summarize and discuss
implications in Sec.~\ref{sec:conc}.

\section{Method}   \label{sec:method}

To create a mock catalog, cluster-sized halos are
extracted from a dark matter only N-body simulation, and then the
gas distribution inside each halo potential is determined.
This section describes the gas model and the construction of the
halo catalog which  is used.

\subsection{The Gas Prescription}   \label{sec:gasp}

A given DM halo is enclosed in a cubic mesh, 
using a cell size, $l$, twice the N-body particle spline softening length.
The DM density, $\rho_{{\rm D}k}$, 
and gravitational potential, $\phi_k$,
are computed for each cell $k$ using a Particle-Mesh code
with non-periodic boundary conditions.
The cell with the lowest potential, $\phi_0=MIN(\phi_k)$, is
considered the center of the cluster.
Particle velocities in the rest frame of the cluster
(defined as the mean velocity of the 125 particles 
closest to the cluster center)
are used to find the KE per unit volume, $\onehalf t_{{\rm D}k}$,
in each cell.
The virial radius 
\citep[from spherical top-hat collapse;][]{BryanN98}
can then be computed,
as well as the halo virial mass, $M_D$, and energy, $E_D$.

It is assumed that gas originally followed the DM inside the virial radius, 
with density $f_c\rho_{{\rm D}k}$ and
KE $f_c \onehalf t_{{\rm D}k}$, where $f_c\equiv \Omega_b/\Omega_m$.
A certain amount of the gas, $M_\star$, will have turned into stars.
We parameterize this mass by the fraction
$f_*\equiv M_*/(f_cM_D)$ at z=0;
the evolution of $M_*$ with redshift is
described in Appendix \ref{sec:starform}.
This stellar mass presumably formed from the portion of the gas with
the lowest entropy and shortest cooling time, so
cells are ranked by binding energy
and then are checked off, starting with the most bound,
until the sum of their gas masses equals $M_\star$;
the gas mass and energy in these cells are set to zero.
Thus the initial mass $M_g$ and energy $E_g$ of the remaining gas are:
\begin{equation} \label{eqn:initmass}
M_g = \sum\limits_{k}  f_c  \rho_{{\rm D}k}  l^3
\hspace{1cm} ,
\end{equation}
\begin{equation} \label{eqn:initegas}
E_g = \sum\limits_{k}  f_c
\left\{ \phi_k\rho_{{\rm D}k}+\onehalf t_{{\rm D}k} \right\} l^3
\hspace{1cm} ,
\end{equation}
summing over all cells inside $r_{vir}$ except
those marked off for star formation.
The surface pressure $P_s$ on the gas at the virial radius
is estimated from the kinetic energy
in a buffer region eleven cells thick:
\begin{equation} \label{eqn:psurf}
P_s = \frac{1}{3}N_{b}^{-1}\sum\limits_{k=1}^{N_{b}} f_c t_{{\rm D}k}
\hspace{1cm} ,
\end{equation}
summing over the $N_{b}$ cells in the buffer region
$r_{vir}<r_k<r_{vir}+11l$.

From this initial state, it is assumed 
the gas is rearranges itself 
into hydrostatic equilibrium in
the DM potential, with
a polytropic equation of state of index $\Gamma=1.2$
\citepalias[as indicated by both observation and simulation;][]{OstrikerBB05}.
The resulting gas pressure $P$ and density $\rho$ are given by
$P_k = P_0 \theta_k^\frac{\Gamma}{\Gamma-1}$ and
$\rho_k = \rho_0 \theta_k^\frac{1}{\Gamma-1}$, 
where the polytropic variable
\begin{equation} \label{eqn:3dtheta}
\theta_k \equiv 1 + \frac{\Gamma-1}{\Gamma(1+\delta_{rel})}
  \frac{\rho_0}{P_0}\left(\phi_0-\phi_k\right)
\hspace{1cm} .
\end{equation}
Here $\delta_{rel}$ is a nonthermal component of pressure, assumed
to be everywhere proportional to thermal pressure; the total is thus
$P_{tot}=(1+\delta_{rel})P$.
The pressure $P_0$ and density $\rho_0$ at the cluster center
are found using two constraints:
requiring conservation of energy, and matching the external
surface pressure.
For a given $P_0$ and $\rho_0$, the final radius
$r_f$ of the gas initially inside $r_{vir}$  is found by summing
outwards from the cluster center until the initial mass $M_g$
is enclosed.
Note this means that gas may expand or contract, changing the gas
fraction inside $r_{vir}$,
and also doing mechanical work.
Assuming surface pressure changes little with radius,
the change in energy is proportional to the change in volume:
$\Delta E_p=(4\pi/3)(r_{vir}^3-r_f^3)P_s$.  The constraint equation for
conservation of energy is thus 
\citepalias[see][]{BodeOWS07}:
\begin{equation} \label{eqn:3dfe}
E_f=\sum\limits_{r_k<r_f} \left\{
\rho_0 \theta_k^\frac{1}{\Gamma-1} \phi_k
+ \frac{3}{2}(1+2\delta_{rel})P_0\theta_k^\frac{\Gamma}{\Gamma-1} \right\}l^3
= E_g + \Delta E_P + \epsilon_D\left| E_D\right| + \epsilon_FM_Fc^2
\hspace{1cm} .
\end{equation}
Here there are two additional inputs to the energy budget.  The
penultimate term
represents energy input to the gas from dynamical processes, 
assumed to be proportional to $E_D$;
the final term
represents feedback energy from collapsed objects, taken to be
proportional to the mass of formed stars, $M_F$
(different from $M_*$ once gas recycling is taken into account---
see Appendix \ref{sec:starform}).
These additional energy inputs
will be described in Sec.~\ref{sec:feedback}.
Finally,
requiring the final surface pressure to match the external pressure yields
the second constraint
\begin{equation} \label{eqn:3dfp}
(1+\delta_{rel})N_{rf}^{-1} \sum\limits_{k=1}^{N_{rf}}
P_0\theta_k^\frac{\Gamma}{\Gamma-1}
= P_s
\hspace{1cm} ,
\end{equation}
summing over the $N_{rf}$ bins in the radial range $r_{f}<r<r_{f}+l$.
(This is a change from \citetalias{BodeOWS07}, which used a wider
zone, thereby giving a lower measurement of surface pressure).

\subsection{Creating the model cluster catalog}  \label{sec:makeclust}

The gas prescription was applied to clusters
selected from a large-scale N-body simulation.  
We will refer the reader to \citetalias{BodeOWS07} and \citet{SehgalTHB07}
for further details, as the numerical methods
used to carry out this simulation are the same as described in those papers.
The cosmological parameters were chosen to be
($\Omega_b, \Omega_m, \Omega_{\Lambda}, h, n_s, \sigma_8$) = 
(0.044, 0.264, 0.736, 0.71, 0.96, 0.80),
consistent with the WMAP 5-year results \citep{KomatsuWMAP08}.
The simulation volume is $1000\hinv$Mpc on a side and contains
N=$1024^3$ particles, making the particle mass 
$6.8\times10^{10}\hinv M_\odot$; the particle spline softening 
length is $16.3\hinv$kpc.  At each simulation step particles
were saved in one octant of a thin spherical shell centered on the origin,
thus building up the matter distribution in a past light cone
covering one eighth of the sky.  All halos with a friends-of-friends
mass of at least $2\times 10^{13}\hinv M_\odot$ (293 particles)  
were selected from this light cone, out to redshift $z=0.2$.
Masses and radii at 
the virial and other overdensities are computed, in particular 
$M_{500}$ and $r_{500}$ at a mean overdensity of 500 times critical.
The amount of
star formation and iron production is determined based
on the cluster redshift,
as described in Appendix \ref{sec:starform}.
The gas prescription was then run on each halo;
there are over 15,000 halos in this catalog  with $M_{500}>10^{13}\hinv M_\odot$.
 
To test that resolution is sufficient, a smaller sample of halos
was selected from a higher resolution simulation.  This has the
same cosmological parameters, but the volume is $320\hinv$Mpc on a side,
and the softening length is $3.2\hinv$kpc;  thus mass resolution is
increased by a factor of 30.5, and spatial resolution by a factor of 5.
No significant difference was seen when comparing results for
lower mass halos ($M_{500}\sim 1-3\times 10^{13} h^{-1}M_\odot$) 
at the two resolutions;
halos more massive than this are difficult to deal with at the
higher resolution because of the large grid required.

Once the density, temperature, and metallicity in each cell are known,
the X-ray luminosity in the 0.5--2 keV band is computed with
the MEKAL model \citep{MeweGO85}.   For the cluster temperature,
the mixT code of \citet{Vikhlinin06} was used to compute the
spectroscopic temperature $T_X$ in the projected annulus $[0.15-1]R_{500}$;
this is the definition of cluster temperature used throughout this paper.

There are a number of parameters which must be specified:
$f_*(M_{vir}), \epsilon_F, \epsilon_D$, and $\delta_{rel}$.
We will calibrate the model by comparing to X-ray data,
in particular by matching the gas fraction as a function of $T_X$.
Two subsamples from are selected from the halo catalog to match
the properties of the two data sets we use for the calibration
(this data is shown as a function of $M_{500}$ in 
Figs.~\ref{fig:sfeffect} and \ref{fig:fbeffect}).
The first data set is from \citet{VikhlininKFJMMVS06}, which consists
mostly of higher $T_X$ clusters; so, for comparison, we select all
halos with $M_{500}>3\times 10^{14}M_\odot$ in the $z\leq 0.2$ light
cone.  There are 76 halos in this subsample, with temperatures
above 4 keV for most parameter choices.
For a given model, we compute a
$\chi_V^2=\sum [f_g(model)-f_g(obs)]^2/\sigma(obs)$,
summing for each observed cluster over all model points
having a $T_X$ within 4-$\sigma$ of the observed $T_X$.
The second observational set is of low redshift groups ($T_X\sim 1-2$keV)
from  \citet{SunVDJFV09}; so, for the second subsample, we select all
model halos with $z<0.0485$ and 
$M_{500}>3\times 10^{13}M_\odot$.  This subsample also contains 76
halos, most with $T_X<3$keV.  A $\chi_S^2$ is computed from this sample
in an analogous manner as above.  When searching for a best fit
model we will minimize $\chi^2=\chi_V^2+0.5\chi_S^2$; the factor of 0.5
ensures that the two terms contribute roughly equally near the minimum
value.

\section{The Role of Feedback and Star Formation}  \label{sec:feedback}

\begin{figure}
\plottwo{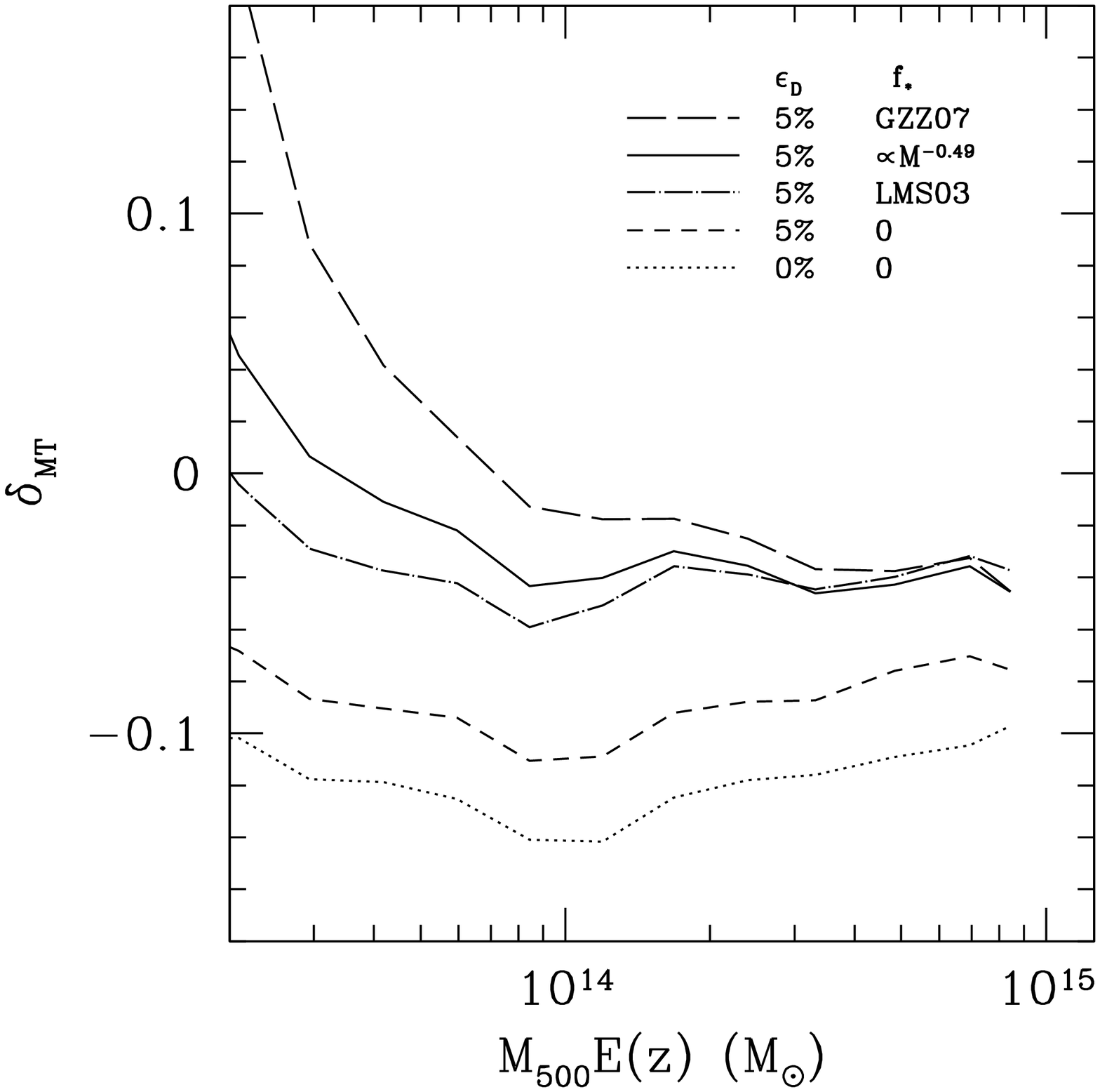}{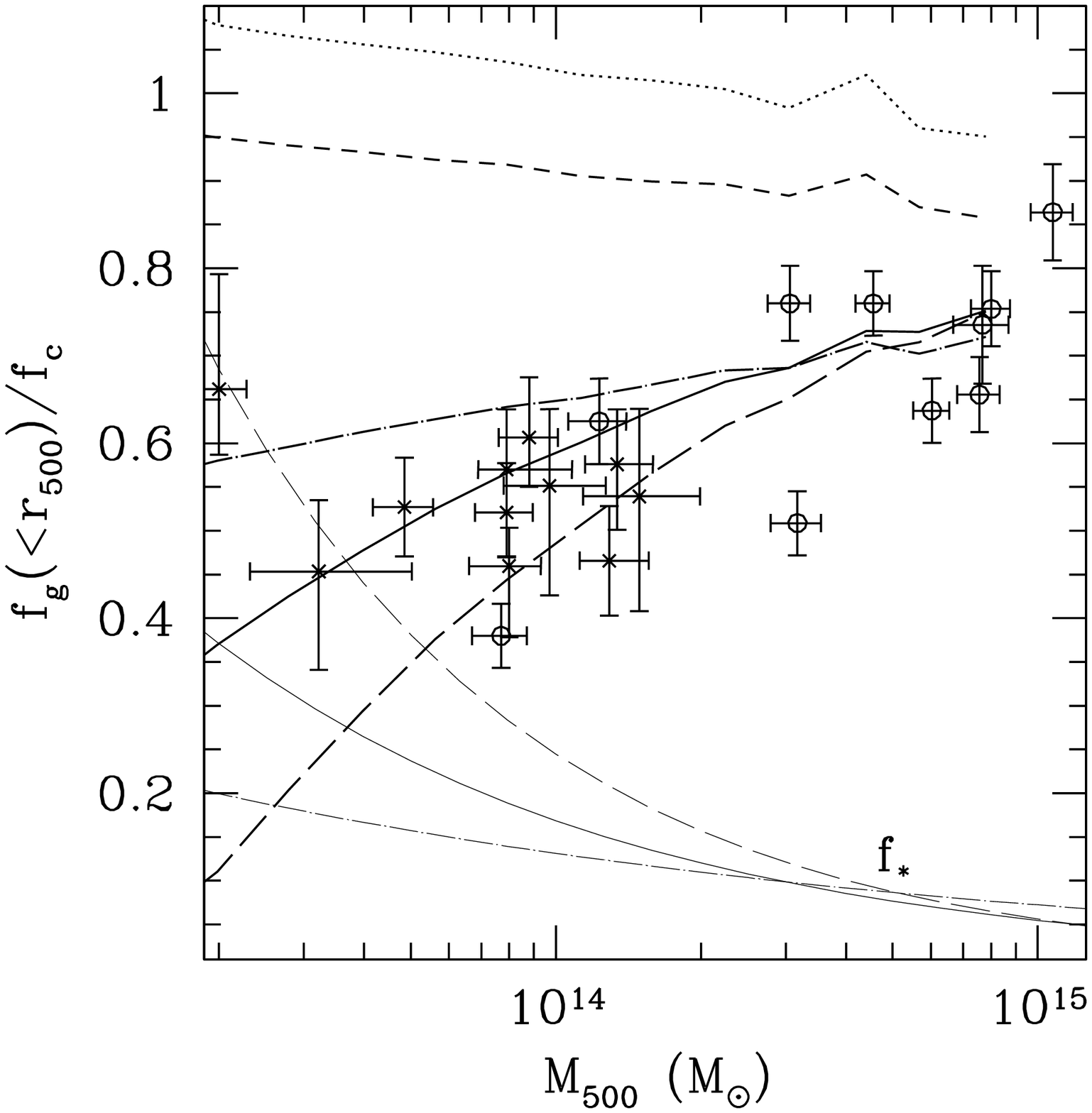}
\caption{Effects of varying star formation.
{\em a)} Fractional difference from the observed
mass-temperature relation of \citet{VikhlininBEFHJKMNQV09}.
{\em b)}  Stellar (thinner lines) and gas (thicker lines)
fractions, normalized to the cosmic average. 
Shown are models with: no star formation
and $\epsilon_D=0$ or $0.05$
(dotted and short-dashed lines, respectively); also
$\epsilon_D=0.05$ and
$f_*\propto M^{-0.26}$ \citep[as in][]{LinMS03}, 
$f_*\propto M^{-0.49}$, and
$f_*\propto M^{-0.64}$ \citep[as in][]{GonzalezZZ07} 
(dot-dashed, solid, and long-dashed lines).
Data points are from \citet[][circles]{VikhlininKFJMMVS06}
and \citet[][crosses]{SunVDJFV09}.
\label{fig:sfeffect} }
\end{figure}

\begin{figure}
\plottwo{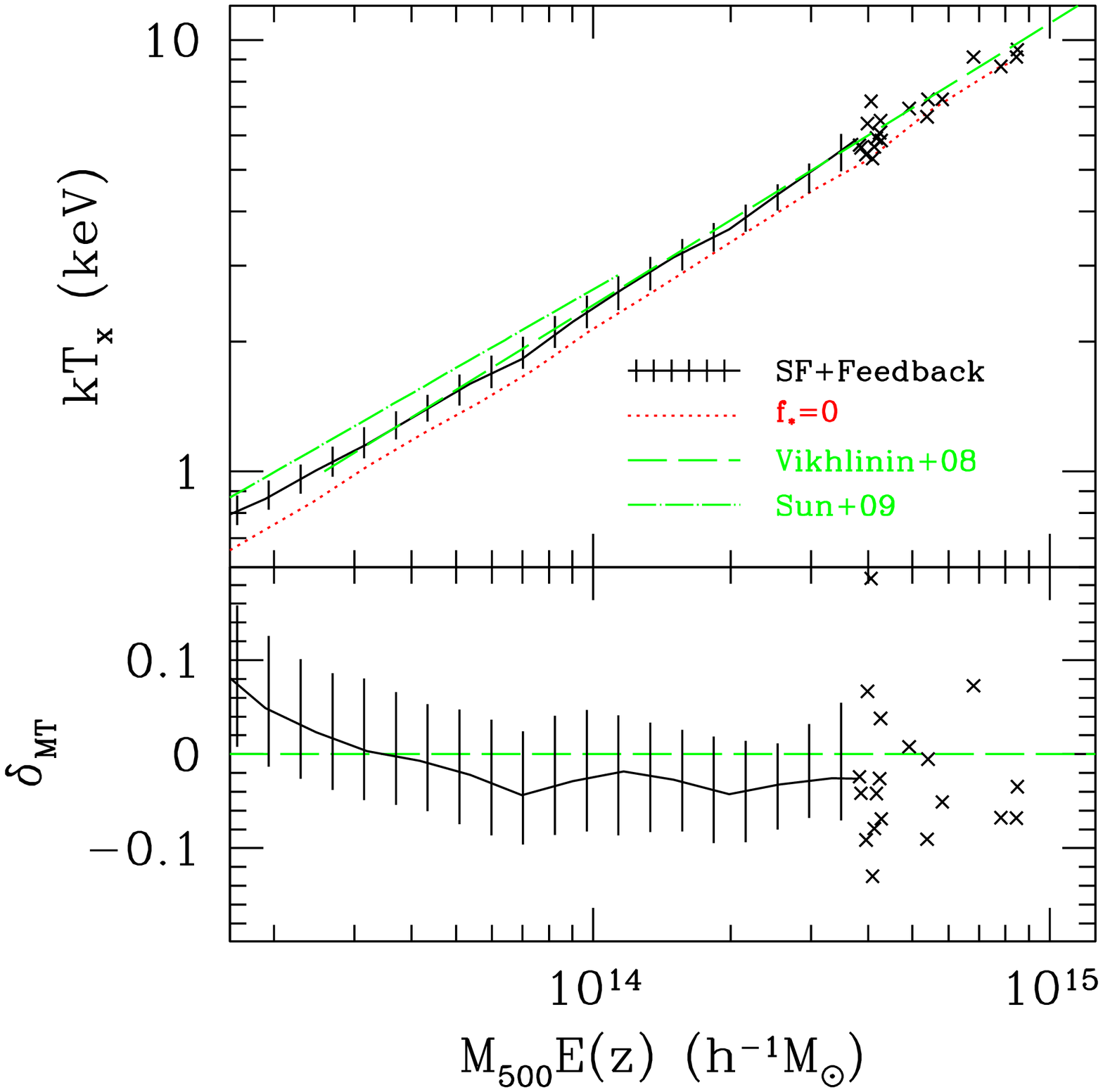}{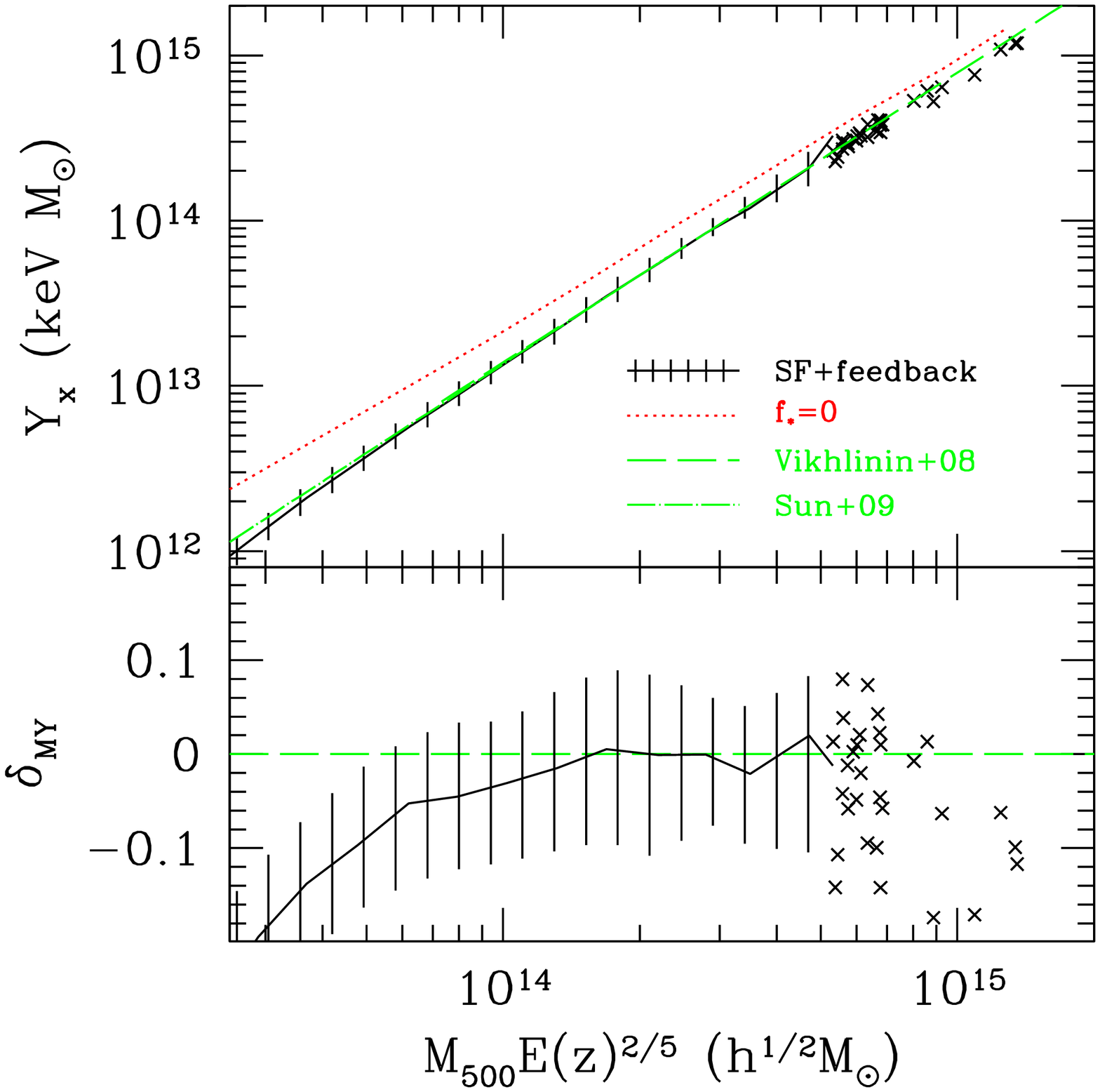}
\caption{Mass-temperature and mass-$Y_X$ relations for
the Standard model with $\epsilon_D=0.05$,
$f_*$ as in \citet{LinMS03}, and
$\epsilon_F=4\times 10^{-6}$. 
Solid lines are the median values, and hash marks
denote the region enclosing 68\% of the halos.
At the high mass end individual halos are shown as $\times$'s.
Dotted lines give the median for the Zero model
(no star formation or feedback).
Straight lines are the observed best-fit power laws:
dashed from \citet{VikhlininBEFHJKMNQV09} and
dot-dashed from \citet{SunVDJFV09}.
Lower panels show the fractional difference from the 
\citet{VikhlininBEFHJKMNQV09} relations.
\label{fig:defmod} }
\end{figure}

It is instructive to first consider the case of
no star formation or energy input, i.e.  $\epsilon_D=f_*=\epsilon_F=0$;
we will refer to this as the ``Zero'' model.
The resulting temperature and $L_X$ are shown as a function of
mass as dotted lines in Fig.~\ref{fig:defmod}.
\citet{VikhlininBEFHJKMNQV09} found the $M-T_X$ relation
is well fit by a power law,
$E(z)M_{500}=M_0(kT/5{\rm keV})^\alpha$, with
$M_0=(3.02\pm0.11)\times10^{14}\hinv M_\odot$ and $\alpha=1.53$.
Fig.~\ref{fig:sfeffect} shows the
fractional difference
$(T_X-T_X(obs))/T_X(obs)$, where $T_X(obs)$ is the value obtained
by plugging $M_{500}$ into the observed relation.
This figure shows that
at a given $M_{500}$, the predicted temperature from this simple
model is lower than observed, by over 10\%.  
At the same time, it shows that the predicted
gas fraction remains near the cosmic mean, and is
independent of mass (quite different from what is observed).
This results in a $Y_X\equiv kT_XM_g(r_{500})$ 
which is too high (by
$\sim 20\%$ at higher masses) and an $M_{500}-Y_X$ relation
with a slope near the self-similar value $Y_X\propto M_{500}^{5/3}$,
slightly shallower than observed.  
Thus we see the simplest set of assumptions yields
an incorrect gas energy.  There are three main processes which can
alter this energy:  transfer of energy from the DM through 
gravitational interactions, star formation, and feedback from
astrophysical sources.  In this section we will consider each
of these in turn.

So far (as in \citetalias{BodeOWS07}), it has been assumed
that the gas originally had the same specific energy as the DM.
However, this ignores an important process during cluster formation:
when mergers occur, energy is transferred from the DM to
the gas.  In fact, the DM can lose up to 10\% of its energy in an
equal mass collision, and
even a 10:1 merger can reduce the DM energy by a
couple of percent \citep{McCarthyBBVPTBLF07}.
Simulations show such mergers are common:
\citet{CohnWhite05} calculate that
a typical cluster will experience four 5:1 accretion
events since $z\sim 2$, and \citet{FakhouriMa08}
find that the mean rate for 10:1 or larger mergers
in clusters is roughly two per halo per unit redshift.
The magnitude of this effect can be estimated from hydrodynamical
simulations, where it has long been found that baryon fraction in
clusters is lower than the universal value 
\citep[e.g.][]{CenO94,EkeNF98,EttoriDBM06,GottloberY07,StanekRE09z}.
For simulations with no star formation or radiative processes,
\citet{CrainEFJMNP07} find that the average
baryon fraction within $r_{200}$ is 90\% of the cosmic mean at $z$=0. 
This result is independent of cluster mass, and at $z$=1 the typical
fraction is only 2 or 3\% higher.
Without star formation and energy feedback, 
the method of Sec.~\ref{sec:gasp} gives a mean 
$f_g(<r_{200})\approx f_c$ for halos 
with $M_{200}\geq 10^{14}h^{-1}M_\odot$
(at low redshift, $z<0.2$). 
Instead assuming that the energy
transferred to the gas is 5\% of the DM energy, i.e. using
$\epsilon_D=0.05$  in Eqn. \ref{eqn:3dfe}, 
this fraction is reduced to $0.92f_c$.
As we are using a lower $\sigma_8$ than \citet{CrainEFJMNP07},
a slightly higher gas fraction at $z=0$ seems appropriate.
Thus we adopt $\epsilon_D=5$\% for the rest of the paper.

\begin{figure}
\plottwo{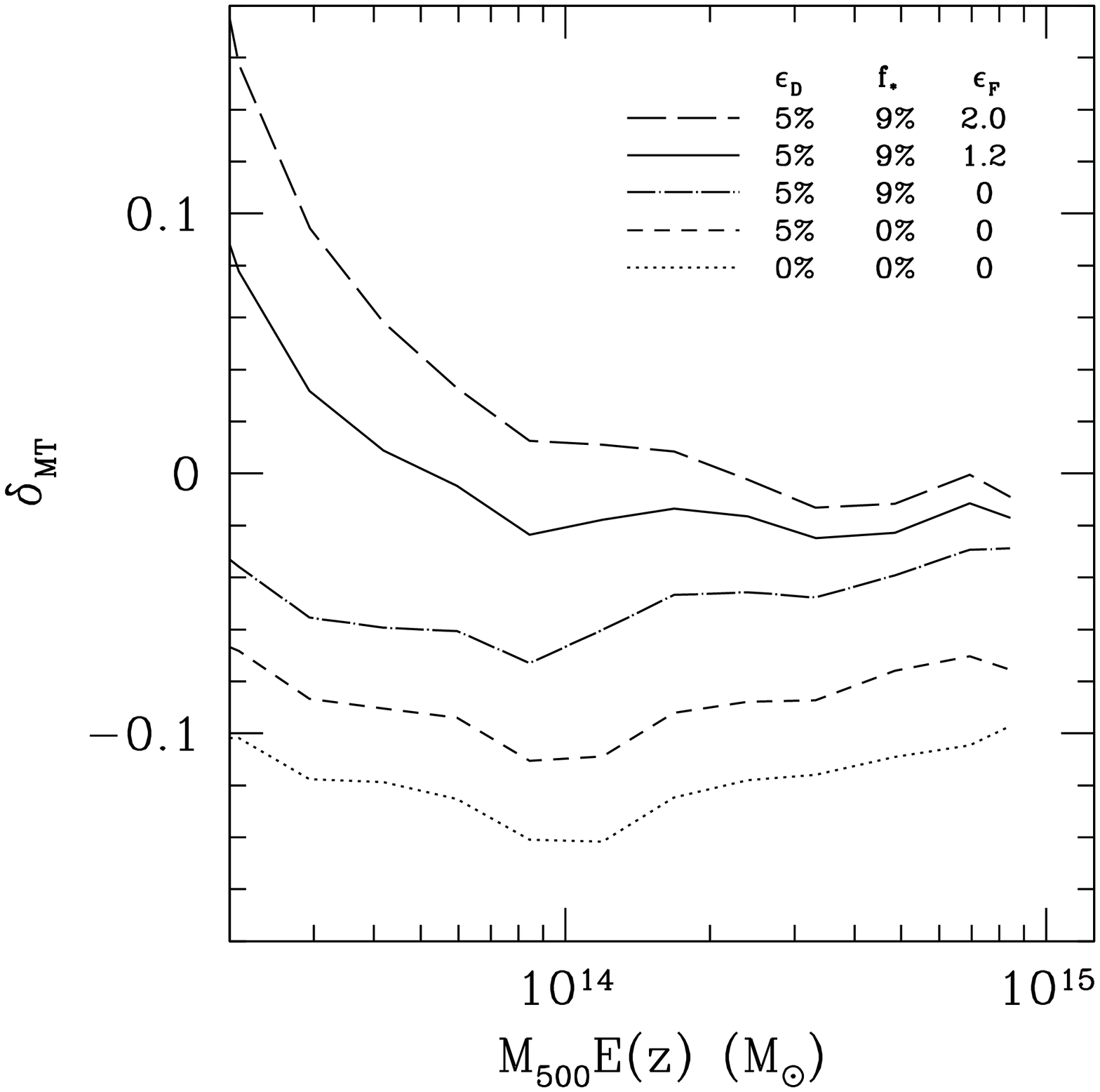}{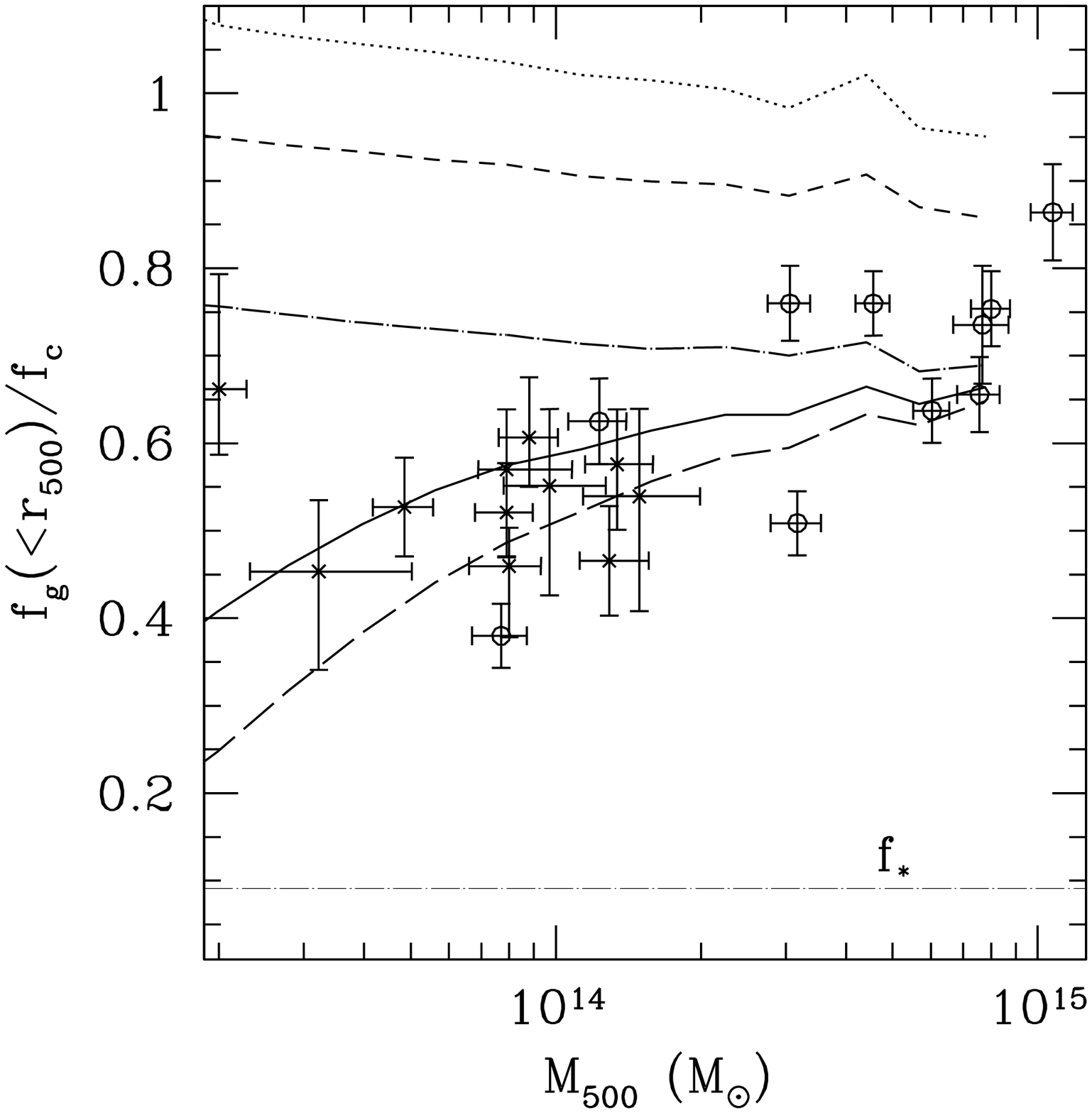}
\caption{Effects of varying feedback.
As in Fig.\ref{fig:sfeffect},
showing models with: no star formation
and $\epsilon_D=0$ or $0.05$
(dotted and short-dashed lines, respectively); also
$\epsilon_D=0.05$ plus
fraction $f_*=9$\% of gas converted to stars, with feedback energies of
$0, 1.2$, and $2.0\times 10^{-5}M_Fc^2$
(dot-long dashed, solid, and long-dashed lines).
\label{fig:fbeffect} }
\end{figure}

Fig.~\ref{fig:sfeffect}, 
and also Fig.~\ref{fig:fbeffect},
display the change in $f_g$ inside $r_{500}$
after including this added energy, as short-dashed lines.  
While lower than before,
it is still higher than observed, and still roughly independent of mass.
The temperature at a given mass increases, but only slightly,
so the $M-Y_X$ relation is still too high and shallow.
Including a term proportional to $E_D$ in this fashion does 
not change the self-similar nature of the model.

Star formation will obviously decrease the gas fraction,
but it has a second effect as well: since it is the 
most bound, lowest entropy gas that is turned into stars,
the remaining gas will have a higher specific energy than
before \citep{VoitBryan01}.
The simplest assumption is to make the efficiency of star
formation independent of cluster mass; 
Fig.~\ref{fig:fbeffect} shows, as dot-dashed lines,
a $\epsilon_D=0.05$
model after also setting $f_*$=9\%
for all halo masses.
The predicted gas fraction drops further---
for massive halos, near the observed value.
Note the change in $f_g$ is more than 9\%; the higher
specific energy of the remaining gas is a significant effect,
causing that gas to expand and further reducing the gas
fraction in the inner parts of the halo.
Temperatures also increase, so at higher masses the
predicted $Y_X$ values are close to those observed.
However, as the method up to this point has been scale-free,
the gas fraction still does not depend on mass; thus, for less
massive clusters, the predicted $f_g$ and $Y_X$ are still too large,
since the slope is still the self-similar value.

There are two ways to change the scale-free nature of the models:
vary star formation efficiency with halo mass, or include feedback energy
proportional to stellar mass.  There is observational support
for both options.
\citet{LinMS03}, with a sample of 27 clusters,
measured the fraction of stellar mass to be
larger for lower mass clusters, following
$M_*(<r_{500})/M_{500}=0.0164(M_{500}/3\times10^{14}M_\odot)^{-\alpha_*}$
with $\alpha_*=0.26\pm 0.09$.
To apply this to our model,
we will assume this relation also holds inside the virial
radius, i.e. $M_*/M_D=M_*(<r_{500})/M_{500}$ at $z=0$.
A model varying $f_*$ with mass in this manner is shown in
Fig.~\ref{fig:sfeffect} as dot-dashed lines.  
As one would expect, this decreases the gas fraction
more in lower mass halos; it also changes the slope
of the $M-T_X$ relation slightly.

The slopes of the $M-T_X$ and $M-Y_X$ relations will clearly
depend on how strongly $M_*$ varies with halo mass.
Recently \citet{GonzalezZZ07} found a much steeper relation
in a sample of 23 clusters:
$M_*(<r_{500})/M_{500}\propto M_{500}^{-0.64}$.
A model where $f_*$ follows this relation is
shown as long-dashed lines in Fig.~\ref{fig:sfeffect}.
The effect is  quite dramatic in low mass clusters
because so much gas is converted into stars ($\sim 50\%$ or more),
and the resulting $T_X$ and
gas fractions tend to overshoot the observed trends at low masses.
If we start with the \citet{LinMS03} and vary the slope,
searching for a best-fit model as described in Sec.~\ref{sec:makeclust},
we find an exponent of $\alpha_*=0.49$ yields a cluster
sample with $M-T_X$ and $M-Y_X$ relations quite close to those 
observed; this model is shown as solid lines in Fig.~\ref{fig:sfeffect}.
We will refer to this choice of parameters
($\epsilon_D=0.05, f_* \propto M_{500}^{-0.49}, \epsilon_F=0,
\delta_{rel}=0$) as our ``SF-only'' model.

The second effect which can break the scale-free nature of the
models is additional energy injection, e.g. from SNe or AGN.
We will assume such feedback
is proportional to the number of stars formed,
$M_F$ (see Appendix \ref{sec:starform}).
As an estimate of the amount of energy under consideration,
for $f_*$=9\% and feedback energy $10^{-5}M_Fc^2$, the energy
added amounts to 0.8 keV per particle initially inside $r_{vir}$.
Core collapse SNe could only add 0.4 keV/particle
(assuming the model in Appendix \ref{sec:starform} and
$10^{51}$ erg per event),
so in this case there must be some additional source 
such as AGN energy from accreting black holes.
If the mass in black holes is $10^{-3}M_*$, and 2\% of the
rest mass energy of the accreted material is converted into
thermal energy \citep{AllenDFTR06}, then 
an additional 1.0 keV/particle
could be added to the gas.

Fig.~\ref{fig:fbeffect} shows the effect of adding feedback,
starting with $\epsilon_D=0.05$ and $f_*=9\%$.
The long-dashed lines are for an extreme model with
$\epsilon_F=2.0\times 10^{-5}$.
For the most massive halos, the feedback energy is not
significant compared to the binding energy of the cluster;
hence there is little change evident.  For less massive
clusters, the added energy is enough to cause the gas to expand, lowering
the gas fraction considerably.  The $M-T_X$ relation becomes
less steep as well, as lower mass halos become hotter;
the $M-Y_X$ relation is thus less steep also,
though in this example the change is so large as to
be incompatible with observations.

Varying the amount of feedback, the best-fit search yields a
more moderate level of $1.2\times 10^{-5}M_Fc^2$,
shown as solid lines in Fig.~\ref{fig:fbeffect}.
We will call this 
($\epsilon_D=0.05, f_*=0.09, \epsilon_F=1.2\times 10^{-5},
\delta_{rel}=0$) the ``Feedback'' model.
Like the SF-only model,
this reproduces quite well the observed $M-T_X$
and $M-Y_X$ relations.
The efficiency $1.2\times 10^{-5}$ is smaller than the equivalent
number used in \citetalias{BodeOWS07} for several reasons.
Part of the difference is merely bookkeeping: \citetalias{BodeOWS07} 
did not consider mass loss from stars, and expressed feedback
as being proportional to $M_*$,  while here we instead use the 
larger mass $M_F$.  More substantially,
\citetalias{BodeOWS07} did not include the 
$\epsilon_D\left| E_D\right|$ term,
instead effectively considering it as part of the feedback.
Finally, here the gas surface pressure is calculated over a 
smaller radial buffer zone than in \citetalias{BodeOWS07}, 
which leads to a slightly lower gas fraction initially.

\begin{figure}
\plotone{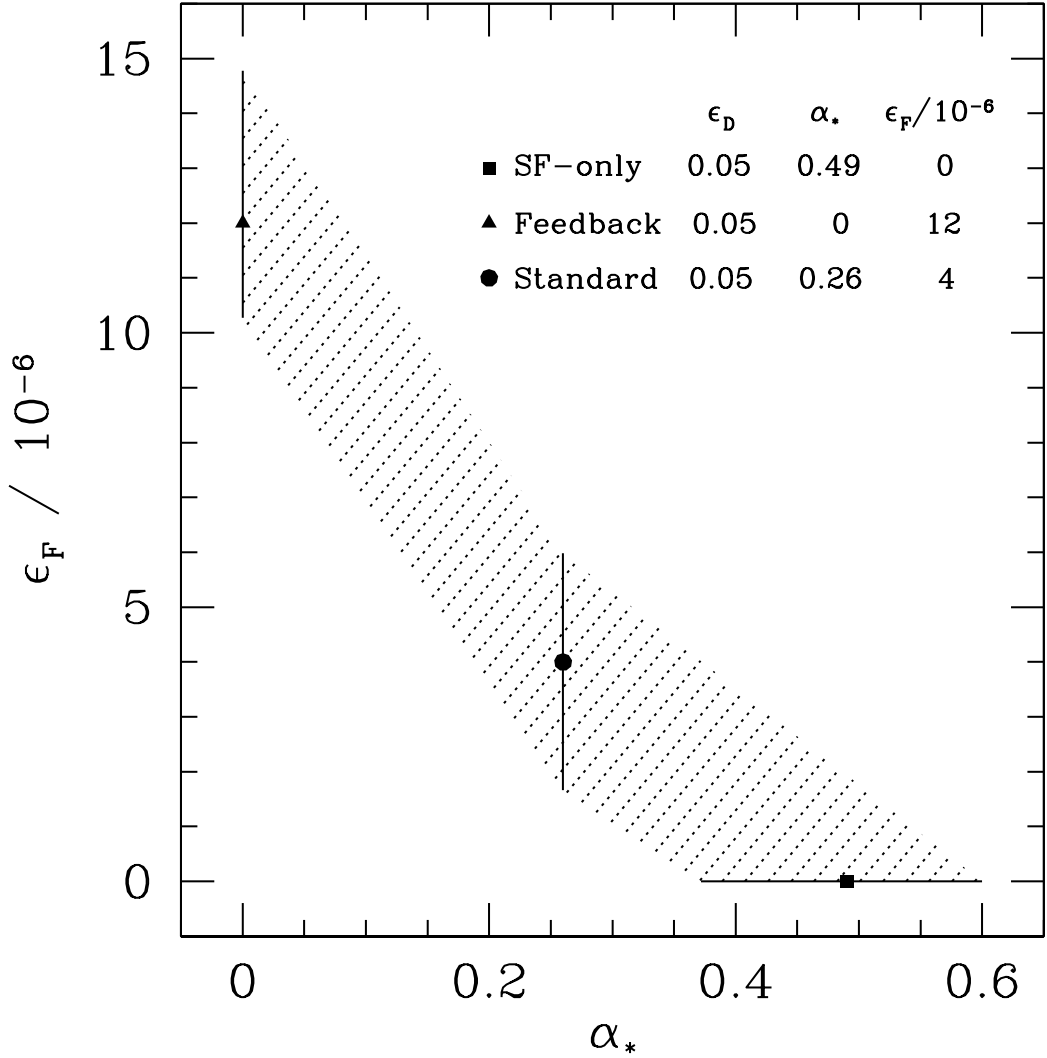}
\caption{Schematic depiction of the degeneracy between varying
star formation efficiency $f_*\propto M_{500}^{-\alpha_*}$ and
feedback energy $\epsilon_F M_Fc^2$.  
Symbols show 
three models based on different assumed $f_*$,  as marked.
The shaded region denotes
a constant $\Delta\chi^2$ from the Standard model, corresponding
to a 99\% confidence interval for two degrees of freedom (this
is only approximate, as
we have not minimized over other input parameters).
\label{fig:eaplane} }
\end{figure}

Figs.~\ref{fig:sfeffect} and \ref{fig:fbeffect} show that
there is a degeneracy between steepening the star formation
efficiency as a function of halo mass and increasing the amount
of feedback per unit stellar mass.  
This is shown schematically in Fig.~\ref{fig:eaplane}.
We have found two
models which both reproduce the X-ray data reasonably well:
the SF-only model, which varies star formation efficiency as 
$f_* \propto M_{500}^{-0.49}$ (steeper than the
values found by \citet{LinMS03} or \citet{Giodiniea09z},
but less steep than \citet{GonzalezZZ07}), and
the Feedback model, which instead has constant $f_*=9\%$ and adds
feedback $1.2\times 10^{-5}M_Fc^2$.
A third option is of course some combination of these two.
Since a stellar mass fraction independent of mass does appear
to be contradicted by observations, we will adopt the star formation 
efficiency as given by \citet{LinMS03} and vary $\epsilon_F$
to find the best-fit model.
This results in a lower level of feedback, $\epsilon_F=4\times 10^{-6}$.
We will call this our ``Standard'' model
($\epsilon_D=0.05, f_* \propto M_{500}^{-0.26}, 
\epsilon_F=4\times 10^{-6}, \delta_{rel}=0$), 
which we will use  for the rest of the paper.
Fig.~\ref{fig:defmod} shows the $M-T_X$
and $M-Y_X$ relations for this third model.
This choice of parameters yields a population 
with thermal energy as a function of mass quite close
to that seen by \citet{VikhlininBEFHJKMNQV09}
and \citet{SunVDJFV09}.
However, keep in mind that there will be a range of models
around this one which will fit the X-ray data equally well--
it would be difficult to distinguish between these three models
based solely on the gas profiles.
The shaded region in Fig. \ref{fig:eaplane} gives a rough
idea of the acceptable range of $\alpha_*$ and  $\epsilon_F$.

\begin{figure}
\plottwo{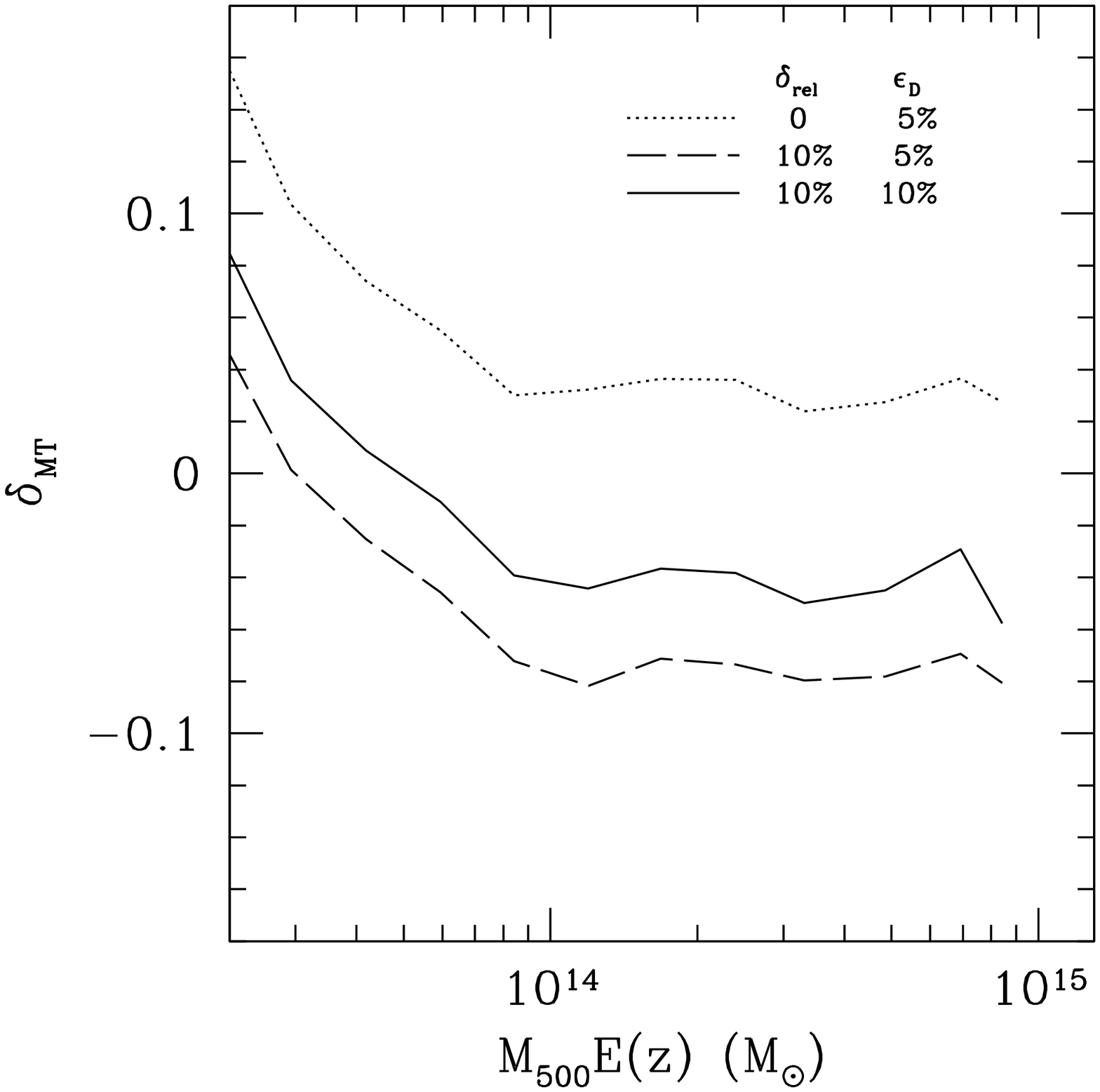}{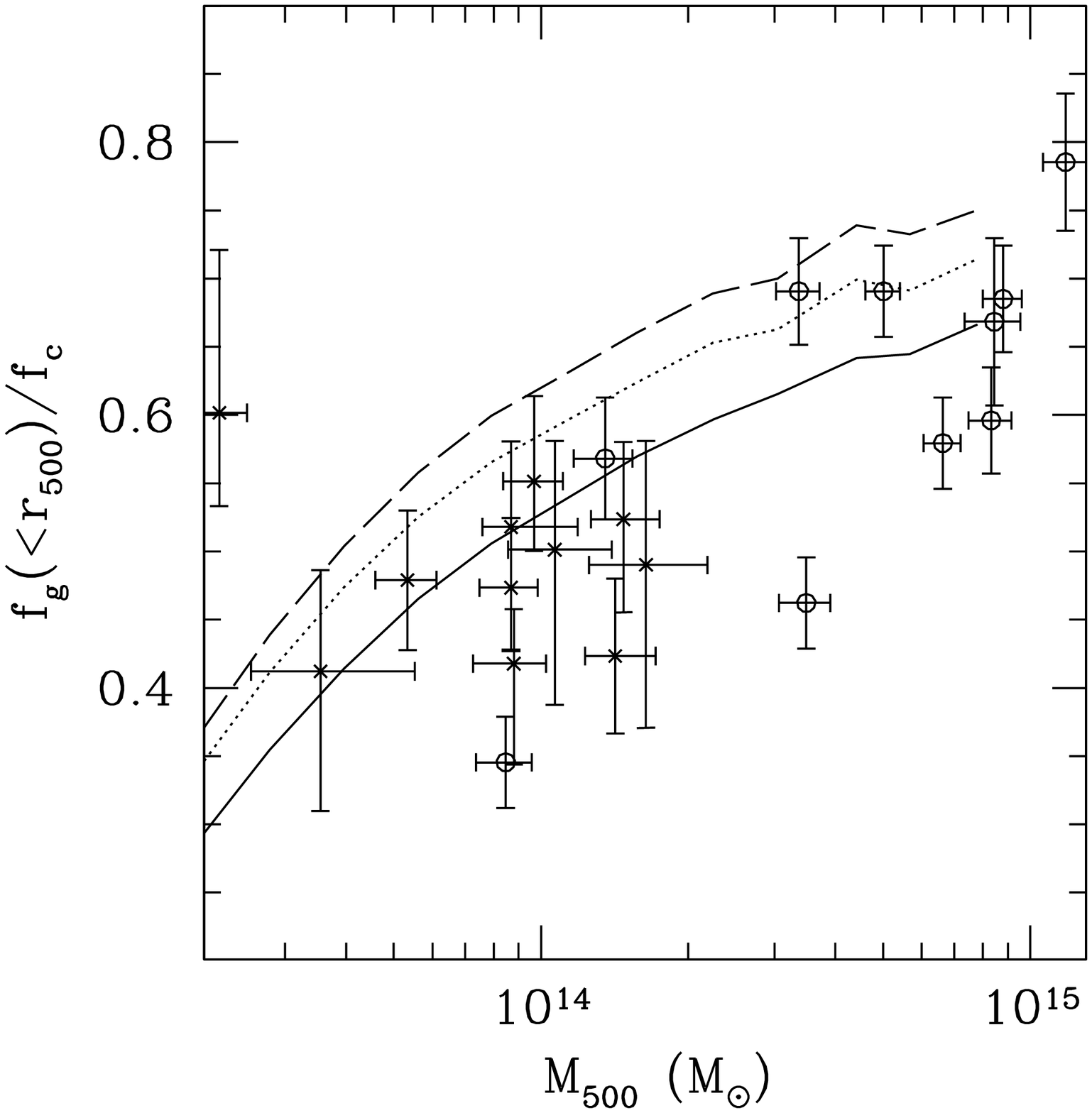}
\caption{Effects of nonthermal pressure.
As in Fig.\ref{fig:sfeffect}, but with observational data
adjusted to assume a 10\% underestimate of cluster mass.
Shown are: the Standard model (dotted lines), after including
$\delta_{rel}=0.10$ (dashed lines), and also doubling
$\epsilon_D$ (solid lines).
\label{fig:dreleffect} }
\end{figure}

There is one further physical effect to be considered, namely
nonthermal pressure support, e.g.
from cosmic rays or magnetic fields \citep[e.g.][]{PfrommerESJD07}.
We will conclude this
section by considering how a moderate level of nonthermal 
support affects our model.
If in fact $\delta_{rel}=0.10$, then the $M_{500}$ derived from
X-rays are too low, because they are derived by
assuming hydrostatic equilibrium.  
Fig.~\ref{fig:dreleffect} shows our Standard model, after adjusting
masses in the observational points by 10\% to account for this;
in  this situation
the model now has too high $T_X$ and $f_g$ at a given $M_{500}$.
Adding $\delta_{rel}=0.10$ to this model means that the gas requires
less thermal pressure to maintain hydrostatic equilibrium, so the
solution to Eqns.~\ref{eqn:3dfe} and \ref{eqn:3dfp} yields gas which
is cooler and correspondingly denser.  This is shown as dashed lines
in Fig.~\ref{fig:dreleffect}; the model is now too cold, and the
gas fraction even higher.  Comparing to Fig.~\ref{fig:sfeffect},
it can be seen that increasing $\delta_{rel}$ has the opposite
effects of increasing $\epsilon_D$.  If we take the $\delta_{rel}=0.10$
model and allow the dynamical energy input to vary, the best-fit result 
has $\epsilon_D=0.10$, shown as solid lines in
Fig.~\ref{fig:dreleffect}.  This puts our predicted clusters back
into agreement with the (adjusted) observations, at the same level
as the Standard model.  We will call this set of parameters
($\epsilon_D=0.10, f_* \propto M_{500}^{-0.26}, 
\epsilon_F=4\times 10^{-6}, \delta_{rel}=0.10$) 
the ``$\delta_{rel}=10\%$'' model.

Increasing $\epsilon_D$ in this way  assumes that non-radiative
hydrodynamic simulations are missing some processes which transfer
energy to gas during cluster formation.
This certainly could be the case.  For example,
the dynamical friction experienced by the stars
assembling into a cD galaxy would add significant amounts of energy
to the ICM \citep{ElZantKK04,Kim07,ConroyO08}.

\section{Radial Profiles}  \label{sec:compother}

Having fixed on a Standard model, in this section
we will examine how well it matches observations
at other radii than $r_{500}$.

Given that our DM halos have the standard density profile,
the $M_{2500}-T_X$ relation of our model should also agree
with observations.
Fitting all halos with $kT_X>1$keV,
the best fit power law relation gives a normalization of
$M_{2500}E(z)=1.25\times 10^{14}h^{-1}M_\odot$
at $kT_X$=5keV.  This agrees
well with observed X-ray samples, using masses derived from
either weak lensing \citep{Hoekstra07} or X-rays
\citep{ArnaudPP05,VikhlininKFJMMVS06}.

The Standard model also has an X-ray luminosity in good agreement 
with the observations of \citet{VikhlininBEFHJKMNQV09},
unlike the Zero model which is a factor of 2-3 more luminous.
This is hardly surprising, since we calibrated the model 
in part on this data.
Above $10^{14}M_\odot$, the predicted
$L_X \propto M_{500}^{1.77}$, 
but below this the relation is
much steeper, approaching $M_{500}^{2.83}$.
\citet{VikhlininBEFHJKMNQV09} find the observed scatter at
a given mass is $\approx \pm 48\%$, while
the model scatter is roughly half of this.
There are many processes occurring in the cores of clusters
which could significantly alter $L_X$ not included in our
model, which would increase the scatter and somewhat alter
the mass dependence.

\begin{figure}
\plotone{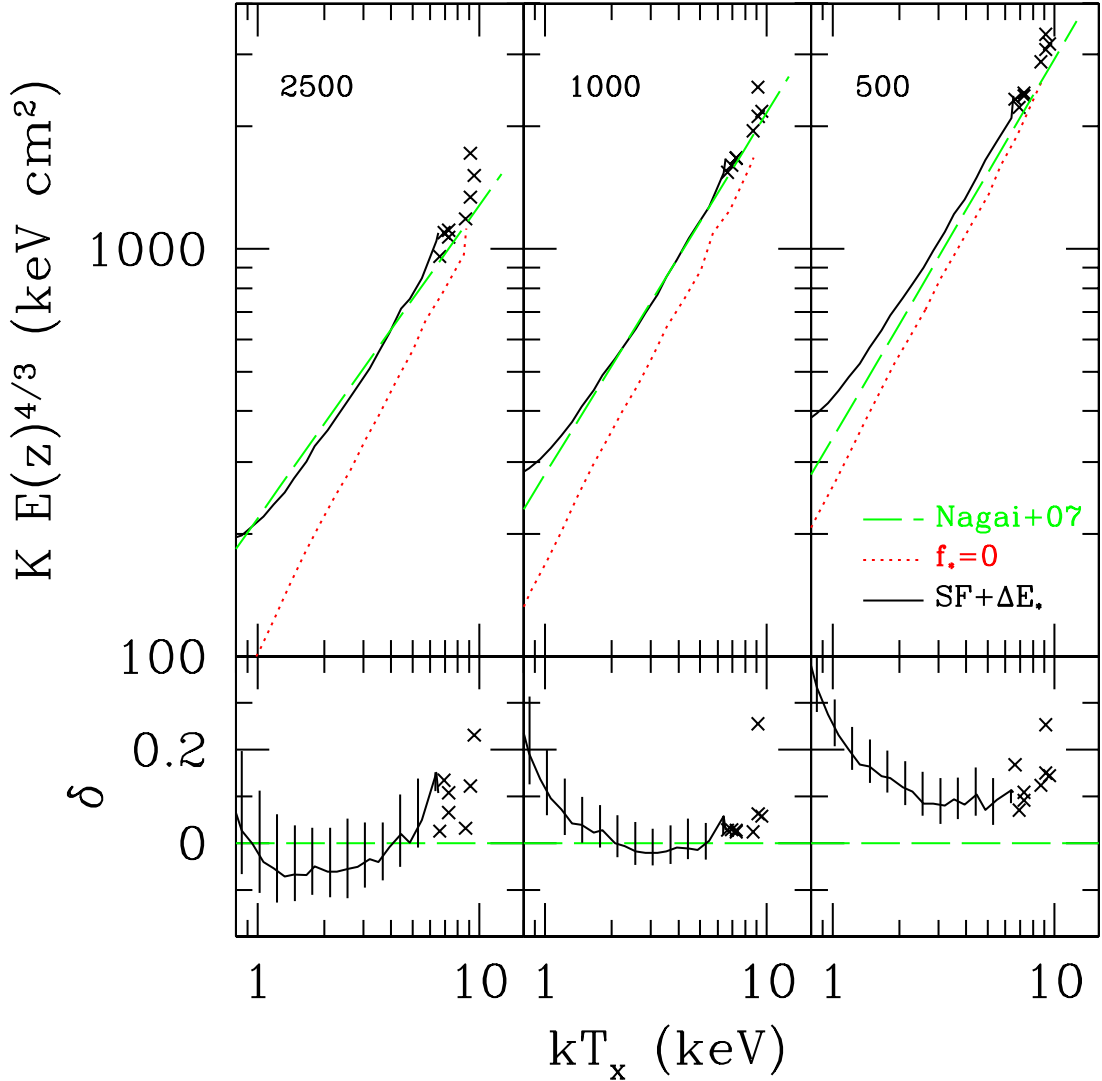}
\caption{Entropy as a function of temperature,
at $r_{2500}$, $r_{1000}$, and $r_{500}$.
Straight dashed lines are the observed best-fit power laws of
\citet{NagaiKV07}.
Solid lines are the median values for the Standard model,
dotted lines for the Zero model.
For the former, clusters at high $T_x$ are shown as individual points.
Lower panels shows the fractional difference of the model from 
the \citet{NagaiKV07} relations.
\label{fig:entdefmod} }
\end{figure}

A fundamental property of the gas is its entropy,
usually defined in cluster studies
as $K\equiv kT/n_e^{2/3}$, where $n_e$ is the electron
number density.
Based on Chandra observations of nearby relaxed clusters,
\citet{NagaiKV07} determined how entropy varies with cluster
temperature at several characteristic
radii, including $r_{500}$, $r_{1000}$, and $r_{2500}$. 
Their best-fit power-law approximations of these data are shown
in Fig.~\ref{fig:entdefmod} as dashed lines, which shows
the entropy at these three radii as a function of $T_X$.
The Zero model gives too low entropy values at all radii;
true clusters have shallower entropy profiles than the
scale-free model predicts.
Once star formation and feedback are included, however,
there is much improved agreement with observed clusters;
the main disagreement is at $r_{500}$, where
the model is roughly 10\% too high.
\citet{SunVDJFV09} also determined $K-T_X$ relations at
these radii, using Chandra archival data and including
many lower temperature groups.
For $T_X>2$keV, the median value from the model clusters
is within 10\% of their power-law fits at $r_{500}$ and $r_{1000}$. 
At $r_{2500}$ the model
is lower than the \citet{SunVDJFV09} relation by roughly 15\%, for
all but the most massive clusters. 
Our model does not consider the history of cluster
formation, nor the fact that the nonadiabatic processes
which can alter the entropy will likely be more pronounced
in the core.

\begin{figure}
\plotone{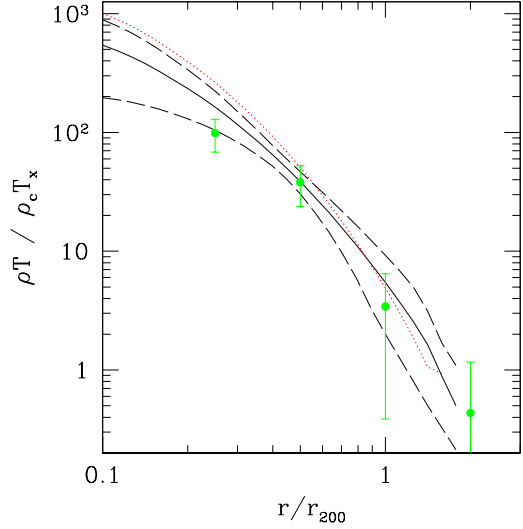}
\caption{The SZ profile as a function of radius.
Solid line is the mean profile for all $kT_x>3$keV clusters;
dashed lines show two standard deviations.
The dotted shows the mean profile for the Zero model.
Points with error bars are the mean profile found from WMAP
data by \citet{AfshordiLNS07}.
\label{fig:naszcomp} }
\end{figure}

\citet{AfshordiLNS07} assembled a heterogeneous catalog of 193
X-ray clusters and then examined {\it WMAP} data to
derive a mean pressure profile using the SZ effect.  
This profile is shown as points
with error bars in Fig.~\ref{fig:naszcomp}; the pressure is
normalized to the critical density times the global cluster $T_X$.
For comparison, we computed the mean profile of the 284 clusters
from our Standard sample with $kT_X>3$keV.  This is shown as solid line in
Fig.~\ref{fig:naszcomp}; the mean profile of the same halos
without star formation or energy input is shown as a dotted line.
To scale the radius, $r_{200}$ is estimated from $kT_X$
\citep[as in Eqn. 1 of][]{AfshordiLNS07}, though using the radius
measured directly from the DM distribution gives the same results.
Outside of $0.5r_{200}$, changes to the gas energy show little
effect on the mean profile.
At smaller radii, star formation and feedback reduce the 
density more than they affect the temperature profile, thus
lowering the pressure profile in the core.  This gives better
agreement with the \citet{AfshordiLNS07} profile, though it
is still higher.  One source of this disagreement is that
they assumed $\rho T\propto r^{-2}$ inside $0.25r_{200}$  when
constraining their profile, while our model is shallower than this;
had they assumed a similarly shallow inner profile, this point
would have been higher.
Note also the redshift and temperature distributions of the model
and observed samples are quite different.

\begin{figure}
\plotone{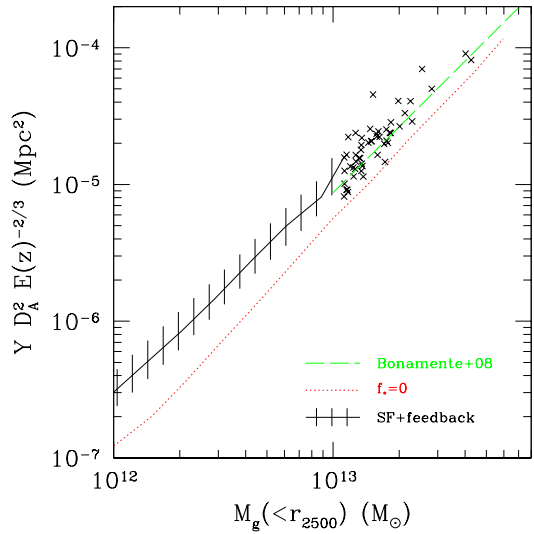}
\caption{The SZ signal as a function of gas mass at overdensity 2500.
The solid line is the median values, with hash marks
denoting the region enclosing 68\% of the halos;
at the high mass end individual halos are shown as $\times$'s.
The dotted line is the median for a model with no star formation
or feedback.
The long-dashed line is the observed relation of \citet{BonamenteJLCNM08}.
\label{fig:bonszcomp} }
\end{figure}

\citet{BonamenteJLCNM08} present a joint SZ and X-ray analysis
of 38 clusters; here  we will compare to their low-redshift
sample of 19 clusters in the range $0.14\leq z\leq 0.30$.
For our Standard sample with $z\leq 0.2$ we calculated the integrated
Compton y-parameter, $Y$ \citep[Eqn. 6 of][]{BonamenteJLCNM08},
out to projected $R_{2500}$;
this is shown in Fig.~\ref{fig:bonszcomp}.
Most of our sample is at lower
gas masses than the clusters used  by \citet{BonamenteJLCNM08},
but where the two samples overlap the agreement is good.
The best fit  power-law found by
\citet{BonamenteJLCNM08}, with slope 1.60, is shown as a
dashed line in Fig.~\ref{fig:bonszcomp}.
A power-law fit to our model halos with 
$M_g(<r_{2500})\geq 8\times 10^{12}\hinv M_\odot$ yields
a slope of 1.61.  
Also shown as a dotted line is the Zero model.
The effect of increasing
the gas energy is to cause gas to expand and reduce the gas mass;
however, this does not change the integrated pressure profile much,
since the gas in the center must still hold up all the gas
above it.
This effect is larger for smaller mass
halos, making the $Y-M_g$ relation less steep.

\begin{figure}
\plotone{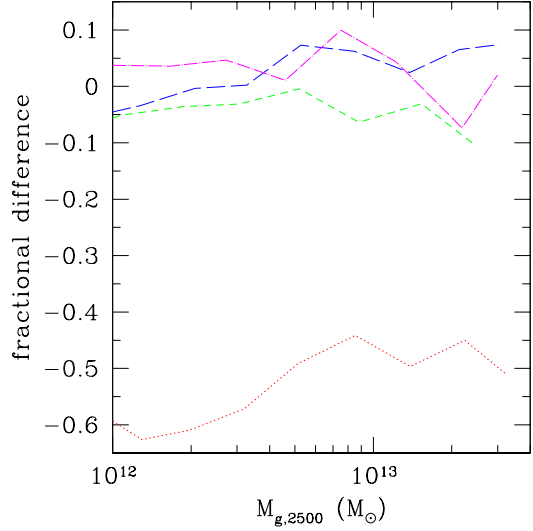}
\caption{Fractional difference of the SZ signal $Y$ from the
Standard model.  Shown are the Zero (dotted line),
SF-only (short-dashed), Feedback (long-dashed),
and $\delta_{rel}=10\%$  (dot-dashed) models.
\label{fig:modszcomp} }
\end{figure}

In Sec.~\ref{sec:feedback} we found several different models which,
given the degeneracy among input parameters, gave similar results.
Since one goal of this is to provide a means  of predicting the SZ
signal from  clusters, it is useful to compare these models.
Fig.~\ref{fig:modszcomp} shows the fractional difference in
the $Y$ value from the Standard model, as a function of
$M_g(<r_{2500})$, for the SF-only, Feedback, and
$\delta_{rel}=10\%$ models.
The median SZ signal is within 10\% of the Standard
model in each case, and they are all roughly a factor of 2 higher
than the Zero model, which does not include star formation and energy injection.
This highlights the importance of normalizing to X-ray data when
modeling the SZ signal.

\section{Summary and Discussion}  \label{sec:conc}

In this paper we have described an improved method for
calculating the gas distribution in a DM potential well,
constraining the model to match the observed X-ray
gas fractions; then we made further
comparisons with X-ray and SZ observations in order
to check, where possible, our conclusions.
Extracting the greatest return from upcoming cluster surveys
will require understanding the state of the intracluster gas.
It is encouraging that the simple model presented here can so
well reproduce observed profiles and trends with cluster mass.
With this model the following main points have demonstrated:
\\
$\bullet$
Simulations show that during mergers the energy of the DM
component will be tapped to provide shock heating to the gas.
Transferring $\sim 5$\% of the binding energy of the DM to
the gas results in gas fractions at large radii which are
in good agreement with non-radiative hydrodynamical simulations.
\\
$\bullet$
Breaking the scale-free nature of the model to bring it into
accord with observed scalings requires either varying the
efficiency of star formation with cluster mass or increasing
the specific energy of the gas by an amount that is not
proportional to the cluster binding energy
(e.g. by being proportional to the stellar mass), 
or both.
\\
$\bullet$
Without including feedback energy,
varying the efficiency of star formation as $f_* \propto M_{500}^{-0.49}$
reproduces observed $M-T_X$ and $M-Y_X$ relations.  This is 
steeper than found by \citet{LinMS03} and \citet{Giodiniea09z},
but less steep than found by \citet{GonzalezZZ07}. 
A relation as steep as $f_* \propto M_{500}^{-0.64}$, as seen by the
latter, would leave too small gas fractions (and that gas at
too high temperatures) in less massive clusters;
however, the difference is less than two sigma.
Also, keep in mind that what is measured is the stellar fraction
inside $r_{500}$, whereas $f_*$ in our model is the fraction inside
$r_{vir}$; the two need not be identical.
We are assuming that
stars have the same radial distribution as DM, but if stars are
instead more centrally located, this would change our predicted
relation.  
\citet{GonzalezZZ07} selected for clusters with a
dominant central galaxy, and for the least massive
clusters $r_{500}$ is in the outskirts of this galaxy.
Because a central dominant galaxy takes up a larger fraction
of the stellar mass in smaller clusters, one could argue that
the stellar fraction inside $r_{500}$ changes more rapidly with
cluster mass than does the fraction inside $r_{vir}$, i.e.
$M_*(<r_{500})$ should vary with halo mass at a larger rate than $f_*$.
\\
$\bullet$
A model with $M_*/M_{500}$ independent of cluster mass but including 
feedback energy proportional to the mass of formed stars can also
reproduce the observed $M-T_X$ and $M-Y_X$.  However, this requires
adding roughly one keV per particle initially inside $r_{vir}$.
This might be possible if the output from accreting black holes 
is converted to thermal energy with high enough efficiency ($\sim 0.02$).
\\
$\bullet$
The most likely possibility is that both 
star formation and feedback are at work.
Assuming $f_*\propto M^{-0.26}$ \citep[as in][]{LinMS03}
and including a lower feedback level of
$4\times 10^{-6}M_Fc^2$
also yields the proper scalings.
Assuming the mass in black holes is $10^{-3}M_*$,   
this energy is $6.5\times 10^{-3}$ of their rest mass
(or 0.33 keV per particle;
note this does not take into account the feedback supplied by
supernovae).  This model was compared to further observations
in Sec.~\ref{sec:compother}, but the other two possibilities
discussed here behave in quite a similar fashion.
\\
$\bullet$
Including nonthermal pressure results in a cooler, denser gas
distribution.  To bring such a model back into agreement with
X-ray observations requires increased energy input.  Once this is
done, the pressure or entropy for a  given $T_X$ is quite similar
to models with $\delta_{rel}=0$, although of course the $M-T_X$
relation is different.

A couple caveats are in order.  
We start with halos drawn from a DM-only simulation;
adding baryons to such a simulation may change the properties
of the halo population \citep{StanekRE09z}.
Also, we have neglected many pertinent
physical processes, such as cooling in the core (note our
definition of cluster temperature excludes the core).  
Because of this, the model halo population likely has too low a
scatter at a given mass.  This is exacerbated by the tight relation
we assume between halo mass and star formation plus feedback.
For example, if we allowed for a variation in $f_*$ at a given
halo mass, clusters with higher $M_*$ would have higher temperatures
and lower gas fractions, and thus lower X-ray luminosities.
This would explain why catalogs of optically selected clusters
tend to be underluminous in X-rays
at the same mass \citep{RykoffMBEJKRSW08}.

\begin{figure}
\plotone{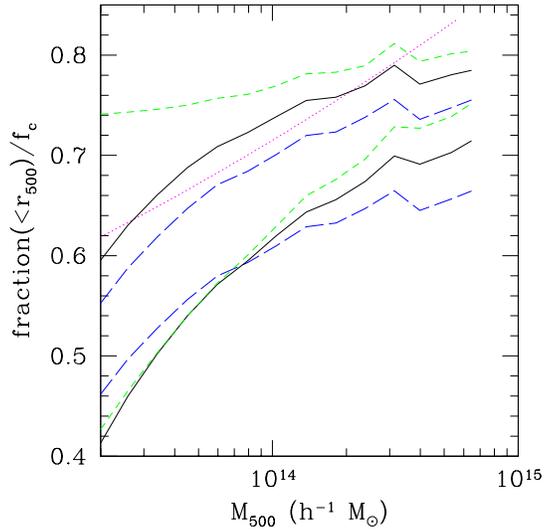}
\caption{Baryon fraction inside $r_{500}$ as a function of mass.
The lower set of lines represents gas fraction, and the upper
set total baryon fraction.
Shown are the Standard (solid lines),
SF-only (short-dashed),
and Feedback (long-dashed) models.
The dotted line is the baryon
fraction (not including intra-cluster light)
determined by \citet{Giodiniea09z}.
\label{fig:baryonf} }
\end{figure}

For most halos in our standard model,
the gas initially assumed to be inside $r_{vir}$ has expanded
outwards.  For hotter ($kT>3$keV) clusters only,
the mean final radius $r_f$ is $1.23\pm 0.13$ (one sigma);
for all clusters the mean $r_f$ is $1.47\pm 0.30$.
A result of this is that for the most massive clusters the
baryon fraction inside $r_{vir}$ is roughly 90\% of the cosmic 
mean, $f_c\equiv \Omega_b/\Omega_m$.
Fig.~\ref{fig:baryonf} shows the 
the baryon fraction inside $r_{500}$ as a function of cluster
mass for differing input  parameters.  The models shown have
similar gas fractions (by design), but varying stellar mass
fractions.  This results in differing total baryon fractions
at lower masses (though, as pointed out above, this may change
if stars are more centrally concentrated than DM).  In every
case, however, the baryon fraction inside $r_{500}$ tops out near
$0.8f_c$ at the highest cluster masses.
Our sample contains four halos with virial
masses above $10^{15}\hinv M_\odot$; their gas fractions within
$r_{200}$ vary from $0.71f_c$ to $0.82f_c$, with a mean of
$0.78f_c$.  This agrees well with the detailed multiwavelength analysis
of four massive clusters by \citet{Umetsuea09z}.
\citet{GonzalezZZ07} also found baryon fractions inside $r_{500}$
near $0.8f_c$ at high masses.  As this work was nearing completion, 
\citet{Giodiniea09z} published a new determination of the baryon
fraction (not including intra-cluster light) as a function of cluster mass.  
Our Standard model is in good agreement with their result,
shown as a dotted line in Fig.~\ref{fig:baryonf}.

The ``missing'' gas in this model is distributed in the outskirts
of clusters.  There it would be difficult to observe directly, being at
low temperatures and densities.  However, \citet{Prokhorov08}
has predicted in this case there would be emission in the
extreme ultraviolet from these outer shells.
Alternatively, as this gas is associated with clusters, its
kinetic SZ signature should be correlated with the cluster
velocity field \citep{HoDS09z,HMH09z}.


\acknowledgments

Computer simulations and analysis were supported by the
National Science Foundation through TeraGrid resources
provided by Pittsburgh Supercomputing Center
and the National Center for Supercomputing Applications
under grant AST070015; computations were also
performed at the TIGRESS high performance
computer center at Princeton University, which is jointly supported by
the Princeton Institute for Computational Science and Engineering and
the Princeton University Office of Information Technology.
PB was partially supported by NSF Grant 0707731.

\appendix

\section{Star formation, gas recycling, and 
metal formation}  \label{sec:starform}

\begin{figure}
\plotone{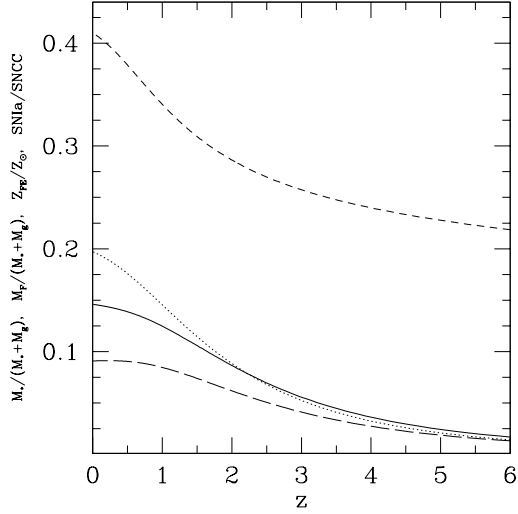}
\caption{
Star formation history as a function of redshift.
{\em Long dashed:} mass in stars, as a fraction of all baryons.
{\em Solid:} total mass of stars formed. 
{\em Dotted:} metallicity with respect to solar.
{\em Short dashed:} the ratio of SNIa to SNCC.
All quantities are measured inside $r_f$, and assume
the redshift zero star/gas ratio is $M_*/M_g$=0.10.
\label{fig:snsum} }
\end{figure}

We will follow
the ``fossil'' model of \citet{NagamineOFC06},
which matches local stellar populations; this
represents a bulge and a disk component with delayed exponentials.
Let $M_F(\tau_j,t)$ denote the mass in stars of type $j$
formed by time $t$, and $F_j=M_F(\tau_j,t_H)$ the total mass 
of such stars 
ever formed by a time $t_H$ corresponding to redshift zero, i.e. today.
(At this point no allowance is made for mass lost from stars.) 
For the time evolution of this stellar mass we will adopt
\begin{equation} \label{eqn:moft}
  M_F(\tau_j,t) = \frac{F_j}{\chi_*(\tau_j,t_H)}\chi_*(\tau_j,t)
  = \frac{F_j}{\chi_*(\tau_j,t_H)}
  \left[ 1 - \left( \totj + 1 \right)\emtotj \right]
\hspace{1cm} .
\end{equation}
This choice means the star formation rate is given by
a delayed exponential
\begin{equation} \label{eqn:sfr}
  \dot{M}_F(\tau_j,t) = 
  \frac{F_j}{\chi_*(\tau_j,t_H)} \dot{\chi}_*(\tau_j,t) = 
  \frac{F_j}{\chi_*(\tau_j,t_H)}\frac{1}{\tau_j}\totj\emtotj
\hspace{1cm} .
\end{equation}
\citet{NagamineOFC06}
found a combination of a bulge population with decay
time $\tau_b$=1.5 Gyr and a disk population with $\tau_d$=4.5 Gyr
is most consistent with local stellar populations
(hence the term fossil).
Since stellar populations in clusters are older than average,
we will assume 90\% of the total stars formed by $z=0$ are
from the bulge component, and the remaining 10\% from the
younger disk component.
The resulting star formation
history is shown in Fig.~\ref{fig:snsum}; there is little
new star formation for $z\la$1.  
This figure assumes $M_*/M_g$=0.1 at $z$=0;
note that if
$r_f$ increases then the amount of gas inside $r_{vir}$ will
decrease, changing the final value of this ratio. 

The supernova rate also depends upon the delay time distribution.
For a burst of star formation at $t$=0, suppose the fraction of
SNe of type $i$ occurring by time $t$ is given by 
$f_{SN}(\tau_i,t)=1-\emtoti$.
Then the number of SNe per unit stellar mass per unit time 
will follow the commonly
used distribution \citep[e.g.][]{Strolger+04,MannucciDVP06}
\begin{equation} \label{eqn:dtd}
  \Phi(\tau_i,t) = \nu_i\dot{f}_{SN}(\tau_i) = \frac{\nu_i}{\tau_i}\emtoti
\hspace{1cm} ,
\end{equation}
where $\nu_i$ is the total number
of SNe of type $i$ per unit mass which will ever occur. 
The SN rate $\dot{S}$, i.e. the number of SNe per unit time, 
is then given by
\begin{equation} \label{eqn:snrate}
  \dot{S}(\tau_i,\tau_j,t) = 
      \int_0^t \Phi(\tau_i,t-s)\frac{dM_F(\tau_j,s)}{ds}ds
  = \frac{ F_j }{ \chi_*(\tau_j,t_H) }  \nu_i \muij^2 
  \left[ \dot{f}_{SN}(\tau_i,t) + 
         \dot{\chi}_*(\tau_j,t) \frac{\tau_j}{\tau_i}
         \left( \frac{1}{\muij} - \frac{\tau_j}{t} 
  \right) \right]
\hspace{1cm} ,
\end{equation}
where $\muij \equiv \tau_i/(\tau_j-\tau_i)$; we will be
assuming $\tau_i \neq \tau_j$ and $\tau_i>0$ throughout.
Finally, the cumulative number of SNe of type $i$ which 
have occurred by time $t$ is  
\begin{eqnarray} \label{eqn:sncum}
  S(\tau_i,\tau_j,t) = \int_0^t \frac{dS(\tau_i,\tau_j,s)}{ds} ds 
  = \frac{ F_j }{ \chi_*(\tau_j,t_H) } \nu_i \chi(\tau_i,\tau_j,t) 
\hspace{1cm} , \\
  \chi(\tau_i,\tau_j,t) = \muij^2 \left[ f_{SN}(\tau_i,t)
    + \frac{1}{\muij}\frac{\tau_j}{\tau_i}\chi_*(\tau_j,t)
    + \frac{\tau_j}{\tau_i}\left( \emtotj - 1 \right) \right]
\hspace{1cm} .
\end{eqnarray}
Thus it only remains to specify $\nu_i$ and $\tau_i$ for
each type of SN.

For core collapse SNe, the delay time is short;
we will adopt $\tau_{CC}=0.01$ Gyr.
We will employ a simple IMF with a slope of -0.5 for
stellar masses in the range $0.1-0.5M_\odot$ and
Salpeter slope -1.35 for $0.5-100M_\odot$ \citep{BaldryG03}.
Assuming all stars in the range 8-50 $M_\odot$
end their lives as SNCC,
then $\nu_{CC}\approx 9.14\times 10^{-3}$.

The picture for SNIa is more complicated.
There is growing evidence for a prompt SNIa component
with a short delay time after star formation
\citep[e.g.][and references therein]{DellaVallePPCMT05,MannucciDVPCCMPT05,MannucciDVP06,Neill+06,Sullivan+06,AubourgJHSS08}.
On the other hand, much observational evidence points towards
much longer delay times of 2-4 Gyr
\citep[e.g.][and references therein]{GallagherGBCJK05,MannucciDVPCCMPT05,Sullivan+06,ForsterSchawinski08z}.
\citet{ScannapiecoBildsten05}
proposed that two evolutionary channels exist, a prompt population
that would be proportional to the recent star formation rate,
and a tardy one with a much longer delay time that would be
proportional to the total stellar mass.
\citet{MannucciDVP06} found that the data are best matched by equal
amounts of SNIa in two populations, one exploding of order $10^8$
yr after star formation and the other following an exponential 
distribution with a decay time of about 3 Gyr.
We will follow this model for SNIa, setting a prompt delay time
$\tau_p=0.05$ Gyr for one half, and a tardy
$\tau_t=3$ Gyr for the other half. 

Assuming SNIa result from 3-8 $M_\odot$ stars,
then $\nu_{i}\approx 0.02754\eta_i$, where $\eta_i$ is
the fraction of the stars in this mass range that go supernova.
This fraction is not well constrained;
\citet{Maoz08} examines a number of different estimates
of this fraction and finds that they vary from 2\% to 40\%,
but also that $\eta\sim15\%$ is consistent with all the estimates
he considered.  Thus we will set 
$\eta_p = \eta_t = 0.07$, or $\nu_p = \nu_t = 0.00193$,
such that there are equal numbers of prompt and tardy SNIa. 

In a $(h,\Omega_m,\Lambda)=(0.7,0.3,0.7)$ cosmology, 
this model gives predicted SN rates per unit mass at $z=0$
of 0.07 SNuM for both SNCC and SNIa.
For comparison, \citet{MannucciMSBDGP08} find
$0.090^{+0.028}_{-0.022}$ and $0.070^{+0.016}_{-0.013}$ SNuM 
($1\sigma$ statistical errors)
for SNCC and SNIa, respectively.
The predicted rates increase with redshift; the predicted SNIa
rate is well within the $1\sigma$ (statistical) errors of the
rate measured by \citet{SharonGMFG07} in the redshift range
0.06--0.19, and of the rate measured by \citet{Graham+08}
in the range 0.2--1.
From the supernova rate one can predict metallicity evolution.
Using iron yields per event of 0.077$M_\odot$ for SNCC and
0.749$M_\odot$ for SNIa,
the resulting history of iron production is also shown in
Fig.~\ref{fig:snsum}.  
Even though there is little star formation at low redshift,
iron production continues because of tardy SNIa.
Note the predicted metallicity will be proportional to 
the amount of star formation; also note that we are
assuming spatial homogeneity.
Fig.~\ref{fig:snsum} also shows the predicted cumulative ratio of
SNIa to SNCC; there are roughly 2.5 SNCC for each SNIa, and SNCC
produce $\sim$20\% of the iron.

\begin{figure}
\plotone{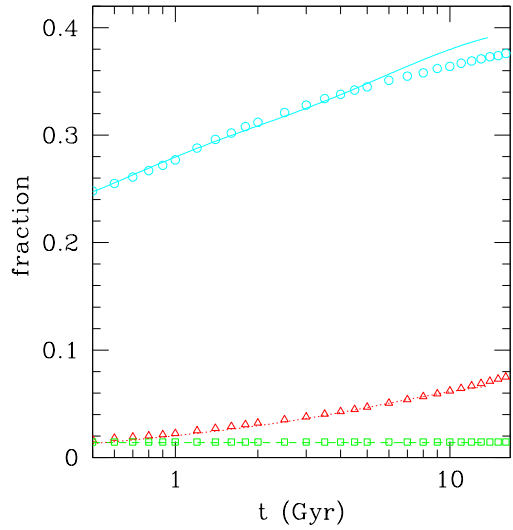}
\caption{Fraction of gas recycled and locked up in remnants as
a function of time for a single starburst at $t=0$.
From top to
bottom, the lines are fraction of recycled gas (solid), in white
dwarfs (dotted), and in black holes or neutron stars (dashed).
For comparison, the same quantities as predicted by the 
PEGASE.2 code \citep{FiocRV99z} are shown as circles, triangles,
and squares, respectively.
\label{fig:remfrac} }
\end{figure}

The rate of change in the actual stellar mass is the difference between
the star formation rate and the rate at which mass is lost from stars.
Codes exist for tracking the mass loss from a stellar
population, e.g. the PEGASE.2 code of \citet{FiocRV99z}.
We will instead assume that all mass loss, whether from SNe
or winds, can be modeled with Eqn.~\ref{eqn:snrate}.  Then
the stellar mass $M_*$ follows
\begin{equation} \label{eqn:dmsdt}
  \dot{M_*}(\tau_j,t) = \dot{M}_F(\tau_j,t) - 
    \sum\limits_{i} y_i \dot{S}(\tau_i,\tau_j,t)
\hspace{1cm} ,
\end{equation}
where the mass yield $y_i$ is the mean initial mass of
stars of type $i$ minus the mean remnant mass, if any.  
The delay times are adjusted to match 
the mass loss calculated by the PEGASE.2
code\footnote{Available at
\url{http://www2.iap.fr/users/fioc/PEGASE.html}}
The choices of delay times and yields are summarized in
Table \ref{tab:starparams}.  
We assume SNCC leave behind
neutron stars of mass 1.4$M_\odot$ and SNIa leave
no remnant.  Stars larger than 50$M_\odot$ are all assumed
to collapse to 2$M_\odot$ black holes, on the same time
scale as SNCC.
Stars in the $5-8M_\odot$ range which do not go supernova
instead remain as 0.55$M_\odot$ white dwarfs.  The 
delay time for mass loss from these stars is set to
0.05 Gyr, slower than the prompt SNIa component;
this choice is required to match the mass loss 
on short ($<1$ Gyr) time scales.
Stars in the $1.1-3M_\odot$ range also become white
dwarfs, with mass loss on a much longer time scale
of 6 Gyr; the lower limit of $1.1M_\odot$ is treated
as a free parameter, set to match the 
mass in white dwarfs at long time scales.
Given these parameters,
the evolution of the recycled gas fraction and the
mass in remnants resulting from a single
starburst is shown in Fig.~\ref{fig:remfrac};
also shown for the same IMF are the predictions
of the PEGASE.2 code.
Our relatively simple scheme follows the more detailed
evolutionary code reasonably well.

\bibliography{}

\begin{thebibliography}{101}\setlength{\itemsep}{-2.01mm}
\expandafter\ifx\csname natexlab\endcsname\relax\def\natexlab#1{#1}\fi


\bibitem[{{Afshordi} {et~al.}(2007){Afshordi}, {Lin}, {Nagai}, \&
  {Sanderson}}]{AfshordiLNS07}
{Afshordi}, N., {Lin}, Y.-T., {Nagai}, D., \& {Sanderson}, A.~J.~R. 2007,
  \mnras, 378, 293

\bibitem[{{Afshordi} {et~al.}(2005){Afshordi}, {Lin}, \&
  {Sanderson}}]{AfshordiLS05}
{Afshordi}, N., {Lin}, Y.-T., \& {Sanderson}, A.~J.~R. 2005, \apj, 629, 1

\bibitem[{{Allen} {et~al.}(2006){Allen}, {Dunn}, {Fabian}, {Taylor}, \&
  {Reynolds}}]{AllenDFTR06}
{Allen}, S.~W., {Dunn}, R.~J.~H., {Fabian}, A.~C., {Taylor}, G.~B., \&
  {Reynolds}, C.~S. 2006, \mnras, 372, 21

\bibitem[{{Arnaud} \& {Evrard}(1999)}]{ArnaudEvrard99}
{Arnaud}, M. \& {Evrard}, A.~E. 1999, \mnras, 305, 631

\bibitem[{{Arnaud} {et~al.}(2005){Arnaud}, {Pointecouteau}, \&
  {Pratt}}]{ArnaudPP05}
{Arnaud}, M., {Pointecouteau}, E., \& {Pratt}, G.~W. 2005, \aap, 441, 893

\bibitem[{{Arnaud} {et~al.}(2007){Arnaud}, {Pointecouteau}, \&
  {Pratt}}]{ArnaudPP07}
---. 2007, \aap, 474, L37

\bibitem[{{Ascasibar} \& {Diego}(2008)}]{AscasibarD08}
{Ascasibar}, Y. \& {Diego}, J.~M. 2008, \mnras, 383, 369

\bibitem[{{Ascasibar} {et~al.}(2003){Ascasibar}, {Yepes}, {M{\"u}ller}, \&
  {Gottl{\"o}ber}}]{AscasibarYMS03}
{Ascasibar}, Y., {Yepes}, G., {M{\"u}ller}, V., \& {Gottl{\"o}ber}, S. 2003,
  \mnras, 346, 731

\bibitem[{{Aubourg} {et~al.}(2007){Aubourg}, {Tojeiro}, {Jimenez}, {Heavens},
  {Strauss}, \& {Spergel}}]{AubourgJHSS08}
{Aubourg}, E., {Tojeiro}, R., {Jimenez}, R., {Heavens}, A.~F., {Strauss},
  M.~A., \& {Spergel}, D.~N. 2007, ArXiv e-prints, 707

\bibitem[{{Babul} {et~al.}(2002){Babul}, {Balogh}, {Lewis}, \&
  {Poole}}]{BabulBLP02}
{Babul}, A., {Balogh}, M.~L., {Lewis}, G.~F., \& {Poole}, G.~B. 2002, \mnras,
  330, 329

\bibitem[{{Baldry} \& {Glazebrook}(2003)}]{BaldryG03}
{Baldry}, I.~K. \& {Glazebrook}, K. 2003, \apj, 593, 258

\bibitem[{{Balogh} {et~al.}(1999){Balogh}, {Babul}, \& {Patton}}]{BaloghBP99}
{Balogh}, M.~L., {Babul}, A., \& {Patton}, D.~R. 1999, \mnras, 307, 463

\bibitem[{{Bhattacharya} {et~al.}(2008){Bhattacharya}, {di Matteo}, \&
  {Kosowsky}}]{BhattacharyaMK08}
{Bhattacharya}, S., {di Matteo}, T., \& {Kosowsky}, A. 2008, \mnras, 389, 34

\bibitem[{{Bode} {et~al.}(2007){Bode}, {Ostriker}, {Weller}, \&
  {Shaw}}]{BodeOWS07}
{Bode}, P., {Ostriker}, J.~P., {Weller}, J., \& {Shaw}, L. 2007, \apj, 663, 139

\bibitem[{{B{\"o}hringer} {et~al.}(2007){B{\"o}hringer}, {Schuecker}, {Pratt},
  {Arnaud}, {Ponman}, {Croston}, {Borgani}, {Bower}, {Briel}, {Collins},
  {Donahue}, {Forman}, {Finoguenov}, {Geller}, {Guzzo}, {Henry}, {Kneissl},
  {Mohr}, {Matsushita}, {Mullis}, {Ohashi}, {Pedersen}, {Pierini}, {Quintana},
  {Raychaudhury}, {Reiprich}, {Romer}, {Rosati}, {Sabirli}, {Temple}, {Viana},
  {Vikhlinin}, {Voit}, \& {Zhang}}]{Bohringerea07}
{B{\"o}hringer}, H. et al. 2007, \aap, 469, 363

\bibitem[{{Bonamente} {et~al.}(2008){Bonamente}, {Joy}, {LaRoque}, {Carlstrom},
  {Nagai}, \& {Marrone}}]{BonamenteJLCNM08}
{Bonamente}, M., {Joy}, M., {LaRoque}, S.~J., {Carlstrom}, J.~E., {Nagai}, D.,
  \& {Marrone}, D.~P. 2008, \apj, 675, 106

\bibitem[{{Borgani} {et~al.}(2008){Borgani}, {Diaferio}, {Dolag}, \&
  {Schindler}}]{BorganiDDS08}
{Borgani}, S., {Diaferio}, A., {Dolag}, K., \& {Schindler}, S. 2008, Space
  Science Reviews, 134, 269

\bibitem[{{Borgani} {et~al.}(2006){Borgani}, {Dolag}, {Murante}, {Cheng},
  {Springel}, {Diaferio}, {Moscardini}, {Tormen}, {Tornatore}, \&
  {Tozzi}}]{BorganDMCSDMTTT06}
{Borgani}, S., {Dolag}, K., {Murante}, G., {Cheng}, L.-M., {Springel}, V.,
  {Diaferio}, A., {Moscardini}, L., {Tormen}, G., {Tornatore}, L., \& {Tozzi},
  P. 2006, \mnras, 367, 1641

\bibitem[{{Bower} {et~al.}(2008){Bower}, {McCarthy}, \& {Benson}}]{BowerMB08}
{Bower}, R.~G., {McCarthy}, I.~G., \& {Benson}, A.~J. 2008, \mnras, 390, 1399

\bibitem[{{Bryan} \& {Norman}(1998)}]{BryanN98}
{Bryan}, G.~L. \& {Norman}, M.~L. 1998, \apj, 495, 80

\bibitem[{{Burenin} {et~al.}(2007){Burenin}, {Vikhlinin}, {Hornstrup},
  {Ebeling}, {Quintana}, \& {Mescheryakov}}]{BureninVHEQM07}
{Burenin}, R.~A., {Vikhlinin}, A., {Hornstrup}, A., {Ebeling}, H., {Quintana},
  H., \& {Mescheryakov}, A. 2007, \apjs, 172, 561

\bibitem[{{Cen} \& {Ostriker}(1994)}]{CenO94}
{Cen}, R. \& {Ostriker}, J.~P. 1994, \apj, 429, 4

\bibitem[{{Ciotti} \& {Pellegrini}(2008)}]{CiottiP08}
{Ciotti}, L. \& {Pellegrini}, S. 2008, \mnras, 387, 902

\bibitem[{{Cohn} \& {White}(2005)}]{CohnWhite05}
{Cohn}, J.~D. \& {White}, M. 2005, Astroparticle Physics, 24, 316

\bibitem[{{Conroy} \& {Ostriker}(2008)}]{ConroyO08}
{Conroy}, C. \& {Ostriker}, J.~P. 2008, \apj, 681, 151

\bibitem[{{Crain} {et~al.}(2007){Crain}, {Eke}, {Frenk}, {Jenkins}, {McCarthy},
  {Navarro}, \& {Pearce}}]{CrainEFJMNP07}
{Crain}, R.~A., {Eke}, V.~R., {Frenk}, C.~S., {Jenkins}, A., {McCarthy}, I.~G.,
  {Navarro}, J.~F., \& {Pearce}, F.~R. 2007, \mnras, 377, 41

\bibitem[{{Dahle}(2007)}]{Dahle07z}
{Dahle}, H. 2007, ArXiv Astrophysics e-prints

\bibitem[{{Dav{\'e}} {et~al.}(2008){Dav{\'e}}, {Oppenheimer}, \&
 {Sivanandam}}]{DaveOS08}
 {Dav{\'e}}, R., {Oppenheimer}, B.~D., \& {Sivanandam}, S. 2008,
 \mnras, 391, 110

\bibitem[{{Della Valle} {et~al.}(2005){Della Valle}, {Panagia}, {Padovani},
  {Cappellaro}, {Mannucci}, \& {Turatto}}]{DellaVallePPCMT05}
{Della Valle}, M., {Panagia}, N., {Padovani}, P., {Cappellaro}, E., {Mannucci},
  F., \& {Turatto}, M. 2005, \apj, 629, 750

\bibitem[{{Dobbs} {et~al.}(2006){Dobbs}, {Halverson}, {Ade}, {Basu}, {Beelen},
  {Bertoldi}, {Cohalan}, {Cho}, {G{\"u}sten}, {Holzapfel}, {Kermish},
  {Kneissl}, {Kov{\'a}cs}, {Kreysa}, {Lanting}, {Lee}, {Lueker}, {Mehl},
  {Menten}, {Muders}, {Nord}, {Plagge}, {Richards}, {Schilke}, {Schwan},
  {Spieler}, {Weiss}, \& {White}}]{Dobbsea06}
{Dobbs}, M. et al. 2006, New Astronomy Review, 50, 960

\bibitem[{{Dos Santos} \& {Dor{\'e}}(2002)}]{DosSantosDore02}
{Dos Santos}, S. \& {Dor{\'e}}, O. 2002, \aap, 383, 450

\bibitem[{{Eisenhardt} {et~al.}(2008){Eisenhardt}, {Brodwin}, {Gonzalez},
  {Stanford}, {Stern}, {Barmby}, {Brown}, {Dawson}, {Dey}, {Doi}, {Galametz},
  {Jannuzi}, {Kochanek}, {Meyers}, {Morokuma}, \& {Moustakas}}]{Eisenhardtea08}
{Eisenhardt}, P.~R.~M. et al. 2008, \apj, 684, 905

\bibitem[{{Eke} {et~al.}(1998){Eke}, {Navarro}, \& {Frenk}}]{EkeNF98}
{Eke}, V.~R., {Navarro}, J.~F., \& {Frenk}, C.~S. 1998, \apj, 503, 569

\bibitem[{{El-Zant} {et~al.}(2004){El-Zant}, {Kim}, \&
  {Kamionkowski}}]{ElZantKK04}
{El-Zant}, A.~A., {Kim}, W.-T., \& {Kamionkowski}, M. 2004, \mnras, 354, 169


\bibitem[{{Ettori} {et~al.}(2006){Ettori}, {Dolag}, {Borgani}, \&
  {Murante}}]{EttoriDBM06}
{Ettori}, S., {Dolag}, K., {Borgani}, S., \& {Murante}, G. 2006, \mnras, 365,
  1021

\bibitem[{{Fakhouri} \& {Ma}(2008)}]{FakhouriMa08}
{Fakhouri}, O. \& {Ma}, C.-P. 2008, \mnras, 386, 577

\bibitem[{{Finoguenov} {et~al.}(2007){Finoguenov}, {Guzzo}, {Hasinger},
  {Scoville}, {Aussel}, {B{\"o}hringer}, {Brusa}, {Capak}, {Cappelluti},
  {Comastri}, {Giodini}, {Griffiths}, {Impey}, {Koekemoer}, {Kneib},
  {Leauthaud}, {Le F{\`e}vre}, {Lilly}, {Mainieri}, {Massey}, {McCracken},
  {Mobasher}, {Murayama}, {Peacock}, {Sakelliou}, {Schinnerer}, {Silverman},
  {Smol{\v c}i{\'c}}, {Taniguchi}, {Tasca}, {Taylor}, {Trump}, \&
  {Zamorani}}]{Finoguenovea07}
{Finoguenov}, A. et al. 2007, \apjs, 172, 182

\bibitem[{{Fioc} \& {Rocca-Volmerange}(1999)}]{FiocRV99z}
{Fioc}, M. \& {Rocca-Volmerange}, B. 1999, ArXiv Astrophysics e-prints

\bibitem[{{Forster} \& {Schawinski}(2008)}]{ForsterSchawinski08z}
{Forster}, F. \& {Schawinski}, K. 2008, ArXiv e-prints, 804

\bibitem[{{Gallagher} {et~al.}(2005){Gallagher}, {Garnavich}, {Berlind},
  {Challis}, {Jha}, \& {Kirshner}}]{GallagherGBCJK05}
{Gallagher}, J.~S., {Garnavich}, P.~M., {Berlind}, P., {Challis}, P., {Jha},
  S., \& {Kirshner}, R.~P. 2005, \apj, 634, 210

\bibitem[{{Giodini} {et~al.}(2009){Giodini}, {Pierini}, {Finoguenov},
{Pratt},
  {Boehringer}, {Leauthaud}, {Guzzo}, {Aussel}, {Bolzonella}, {Capak},
  {Elvis}
,
  {Hasinger}, {Ilbert}, {Kartaltepe}, {Koekemoer}, {Lilly},
  {McCracken},
  {Salvato}, {Sanders}, {Scoville}, {Sasaki}, {Smolcic}, {Taniguchi},
  {Thompson}, \& {the COSMOS collaboration}}]{Giodiniea09z}
{Giodini}, S. et al.  2009, ArXiv e-prints


\bibitem[{{Gonzalez} {et~al.}(2007){Gonzalez}, {Zaritsky}, \&
  {Zabludoff}}]{GonzalezZZ07}
{Gonzalez}, A.~H., {Zaritsky}, D., \& {Zabludoff}, A.~I. 2007, \apj, 666, 147

\bibitem[{{Gottl{\"o}ber} \& {Yepes}(2007)}]{GottloberY07}
{Gottl{\"o}ber}, S. \& {Yepes}, G. 2007, \apj, 664, 117

\bibitem[{{Graham} {et~al.}(2008){Graham}, {Pritchet}, {Sullivan}, {Gwyn},
  {Neill}, {Hsiao}, {Astier}, {Balam}, {Balland}, {Basa}, {Carlberg}, {Conley},
  {Fouchez}, {Guy}, {Hardin}, {Hook}, {Howell}, {Pain}, {Perrett}, {Regnault},
  {Baumont}, {LeDu}, {Lidman}, {Perlmutter}, {Ripoche}, {Suzuki}, {Walker}, \&
  {Zhang}}]{Graham+08}
{Graham}, M.~L. et al. 2008, \aj, 135, 1343

\bibitem[{{Hernandez-Monteagudo} \& {Ho}(2009)}]{HMH09z}
{Hernandez-Monteagudo}, C. \& {Ho}, S. 2009, ArXiv e-prints

\bibitem[{{Ho} {et~al.}(2009){Ho}, {Dedeo}, \& {Spergel}}]{HoDS09z}
{Ho}, S., {Dedeo}, S., \& {Spergel}, D. 2009, ArXiv e-prints

\bibitem[{{Hoekstra}(2007)}]{Hoekstra07}
{Hoekstra}, H. 2007, \mnras, 379, 317

\bibitem[{{Kaiser}(1991)}]{Kaiser91}
{Kaiser}, N. 1991, \apj, 383, 104

\bibitem[{{Kay} {et~al.}(2007){Kay}, {da Silva}, {Aghanim}, {Blanchard},
  {Liddle}, {Puget}, {Sadat}, \& {Thomas}}]{KaydSABLPST07}
{Kay}, S.~T., {da Silva}, A.~C., {Aghanim}, N., {Blanchard}, A., {Liddle},
  A.~R., {Puget}, J.-L., {Sadat}, R., \& {Thomas}, P.~A. 2007, \mnras, 377, 317

\bibitem[{{Kim}(2007)}]{Kim07} {Kim}, W.-T. 2007, \apjl, 667, L5

\bibitem[{{Koester} {et~al.}(2007){Koester}, {McKay}, {Annis}, {Wechsler},
  {Evrard}, {Bleem}, {Becker}, {Johnston}, {Sheldon}, {Nichol}, {Miller},
  {Scranton}, {Bahcall}, {Barentine}, {Brewington}, {Brinkmann}, {Harvanek},
  {Kleinman}, {Krzesinski}, {Long}, {Nitta}, {Schneider}, {Sneddin}, {Voges},
  \& {York}}]{Koesterea07}
{Koester}, B.~P. et al. 2007, \apj, 660, 239

\bibitem[{{Komatsu} {et~al.}(2008){Komatsu}, {Dunkley}, {Nolta}, {Bennett},
  {Gold}, {Hinshaw}, {Jarosik}, {Larson}, {Limon}, {Page}, {Spergel},
  {Halpern}, {Hill}, {Kogut}, {Meyer}, {Tucker}, {Weiland}, {Wollack}, \&
  {Wright}}]{KomatsuWMAP08}
{Komatsu}, E. et al.  2008, ArXiv e-prints

\bibitem[{{Komatsu} \& {Seljak}(2001)}]{KomatsuSeljak01}
{Komatsu}, E. \& {Seljak}, U. 2001, \mnras, 327, 1353

\bibitem[{{Kosowsky}(2006)}]{Kosowsky06}
{Kosowsky}, A. 2006, New Astronomy Review, 50, 969

\bibitem[{{Lapi} {et~al.}(2005){Lapi}, {Cavaliere}, \& {Menci}}]{LapiCM05}
{Lapi}, A., {Cavaliere}, A., \& {Menci}, N. 2005, \apj, 619, 60

\bibitem[{{Lin} {et~al.}(2008){Lin}, {Wu}, {Umetsu}, {Kock}, {Liu}, {Nishioka},
  {Huang}, {Liao}, {Wang}, \& {Ho}}]{LinWUKLNHLWH08}
{Lin}, K.-Y. et al.
  2008, in Astronomical Society of the Pacific Conference Series, Vol. 399,
  Astronomical Society of the Pacific Conference Series, ed. T.~{Kodama},
  T.~{Yamada}, \& K.~{Aoki}, 384-

\bibitem[{{Lin} {et~al.}(2003){Lin}, {Mohr}, \& {Stanford}}]{LinMS03}
{Lin}, Y.-T., {Mohr}, J.~J., \& {Stanford}, S.~A. 2003, \apj, 591, 749

\bibitem[{{Loewenstein}(2000)}]{Loewenstein00}
{Loewenstein}, M. 2000, \apj, 532, 17

\bibitem[{{Lopes} {et~al.}(2009){Lopes}, {de Carvalho}, {Kohl-Moreira}, \&
  {Jones}}]{LopesCKJ09}
{Lopes}, P.~A.~A., {de Carvalho}, R.~R., {Kohl-Moreira}, J.~L., \& {Jones}, C.
  2009, \mnras, 392, 135

\bibitem[{{Malte Sch{\"a}fer} \& {Bartelmann}(2007)}]{MalteSchaferB07}
{Malte Sch{\"a}fer}, B. \& {Bartelmann}, M. 2007, \mnras, 377, 253

\bibitem[{{Mannucci} {et~al.}(2006){Mannucci}, {Della Valle}, \&
  {Panagia}}]{MannucciDVP06}
{Mannucci}, F., {Della Valle}, M., \& {Panagia}, N. 2006, \mnras, 370, 773

\bibitem[{{Mannucci} {et~al.}(2005){Mannucci}, {Della Valle}, {Panagia},
  {Cappellaro}, {Cresci}, {Maiolino}, {Petrosian}, \&
  {Turatto}}]{MannucciDVPCCMPT05}
{Mannucci}, F., {Della Valle}, M., {Panagia}, N., {Cappellaro}, E., {Cresci},
  G., {Maiolino}, R., {Petrosian}, A., \& {Turatto}, M. 2005, \aap, 433, 807

\bibitem[{{Mannucci} {et~al.}(2008){Mannucci}, {Maoz}, {Sharon}, {Botticella},
  {Della Valle}, {Gal-Yam}, \& {Panagia}}]{MannucciMSBDGP08}
{Mannucci}, F., {Maoz}, D., {Sharon}, K., {Botticella}, M.~T., {Della Valle},
  M., {Gal-Yam}, A., \& {Panagia}, N. 2008, \mnras, 383, 1121

\bibitem[{{Maoz}(2008)}]{Maoz08}
{Maoz}, D. 2008, \mnras, 384, 267

\bibitem[{{McCarthy} {et~al.}(2007){McCarthy}, {Bower}, {Balogh}, {Voit},
  {Pearce}, {Theuns}, {Babul}, {Lacey}, \& {Frenk}}]{McCarthyBBVPTBLF07}
{McCarthy}, I.~G., {Bower}, R.~G., {Balogh}, M.~L., {Voit}, G.~M., {Pearce},
  F.~R., {Theuns}, T., {Babul}, A., {Lacey}, C.~G., \& {Frenk}, C.~S. 2007,
  \mnras, 376, 497

\bibitem[{{Mewe} {et~al.}(1985){Mewe}, {Gronenschild}, \& {van den
  Oord}}]{MeweGO85}
{Mewe}, R., {Gronenschild}, E.~H.~B.~M., \& {van den Oord}, G.~H.~J. 1985,
  \aaps, 62, 197

\bibitem[{{Moodley} {et~al.}(2008){Moodley}, {Warne}, {Goheer}, \&
  {Trac}}]{MoodleyWGT08z}
{Moodley}, K., {Warne}, R., {Goheer}, N., \& {Trac}, H. 2008, ArXiv e-prints

\bibitem[{{Muanwong} {et~al.}(2006){Muanwong}, {Kay}, \&
  {Thomas}}]{MuanwongKT06}
{Muanwong}, O., {Kay}, S.~T., \& {Thomas}, P.~A. 2006, \apj, 649, 640

\bibitem[{{Muchovej} {et~al.}(2007){Muchovej}, {Mroczkowski}, {Carlstrom},
  {Cartwright}, {Greer}, {Hennessy}, {Loh}, {Pryke}, {Reddall}, {Runyan},
  {Sharp}, {Hawkins}, {Lamb}, {Woody}, {Joy}, {Leitch}, \&
  {Miller}}]{Muchovejea07}
{Muchovej}, S. et  al. 2007, \apj, 663, 708

\bibitem[{{Nagai} {et~al.}(2007){Nagai}, {Kravtsov}, \&
  {Vikhlinin}}]{NagaiKV07}
{Nagai}, D., {Kravtsov}, A.~V., \& {Vikhlinin}, A. 2007, \apj, 668, 1

\bibitem[{{Nagamine} {et~al.}(2006){Nagamine}, {Ostriker}, {Fukugita}, \&
  {Cen}}]{NagamineOFC06}
{Nagamine}, K., {Ostriker}, J.~P., {Fukugita}, M., \& {Cen}, R. 2006, \apj,
  653, 881

\bibitem[{{Neill} {et~al.}(2006){Neill}, {Sullivan}, {Balam}, {Pritchet},
  {Howell}, {Perrett}, {Astier}, {Aubourg}, {Basa}, {Carlberg}, {Conley},
  {Fabbro}, {Fouchez}, {Guy}, {Hook}, {Pain}, {Palanque-Delabrouille},
  {Regnault}, {Rich}, {Taillet}, {Aldering}, {Antilogus}, {Arsenijevic},
  {Balland}, {Baumont}, {Bronder}, {Ellis}, {Filiol}, {Gon{\c c}alves},
  {Hardin}, {Kowalski}, {Lidman}, {Lusset}, {Mouchet}, {Mourao}, {Perlmutter},
  {Ripoche}, {Schlegel}, \& {Tao}}]{Neill+06}
{Neill}, J.~D. et al. 2006, \aj, 132, 1126

\bibitem[{{Ostriker} {et~al.}(2005){Ostriker}, {Bode}, \&
  {Babul}}]{OstrikerBB05}
{Ostriker}, J.~P., {Bode}, P., \& {Babul}, A. 2005, \apj, 634, 964

\bibitem[{{Pacaud} {et~al.}(2007){Pacaud}, {Pierre}, {Adami}, {Altieri},
  {Andreon}, {Chiappetti}, {Detal}, {Duc}, {Galaz}, {Gueguen}, {Le F{\`e}vre},
  {Hertling}, {Libbrecht}, {Melin}, {Ponman}, {Quintana}, {Refregier},
  {Sprimont}, {Surdej}, {Valtchanov}, {Willis}, {Alloin}, {Birkinshaw},
  {Bremer}, {Garcet}, {Jean}, {Jones}, {Le F{\`e}vre}, {Maccagni}, {Mazure},
  {Proust}, {R{\"o}ttgering}, \& {Trinchieri}}]{Pacaudea07}
{Pacaud}, F. et al. 2007, \mnras, 382, 1289

\bibitem[{{Pfrommer} {et~al.}(2007){Pfrommer}, {En{\ss}lin}, {Springel},
  {Jubelgas}, \& {Dolag}}]{PfrommerESJD07}
{Pfrommer}, C., {En{\ss}lin}, T.~A., {Springel}, V., {Jubelgas}, M., \&
  {Dolag}, K. 2007, \mnras, 378, 385

\bibitem[{{Pratt} {et~al.}(2008){Pratt}, {Croston}, {Arnaud}, \&
  {Boehringer}}]{PrattCAB08z}
{Pratt}, G.~W., {Croston}, J.~H., {Arnaud}, M., \& {Boehringer}, H. 2008, ArXiv
  e-prints

\bibitem[{{Prokhorov}(2008)}]{Prokhorov08}
{Prokhorov}, D.~A. 2008, \aap, 492, 651

\bibitem[{{Puchwein} {et~al.}(2008){Puchwein}, {Sijacki}, \&
  {Springel}}]{PuchweinSS08}
{Puchwein}, E., {Sijacki}, D., \& {Springel}, V. 2008, \apjl, 687, L53

\bibitem[{{Rykoff} {et~al.}(2008){Rykoff}, {McKay}, {Becker}, {Evrard},
  {Johnston}, {Koester}, {Rozo}, {Sheldon}, \& {Wechsler}}]{RykoffMBEJKRSW08}
{Rykoff}, E.~S., {McKay}, T.~A., {Becker}, M.~R., {Evrard}, A., {Johnston},
  D.~E., {Koester}, B.~P., {Rozo}, E., {Sheldon}, E.~S., \& {Wechsler}, R.~H.
  2008, \apj, 675, 1106

\bibitem[{{Scannapieco} \& {Bildsten}(2005)}]{ScannapiecoBildsten05}
{Scannapieco}, E. \& {Bildsten}, L. 2005, \apjl, 629, L85

\bibitem[{{Sehgal} {et~al.}(2007){Sehgal}, {Trac}, {Huffenberger}, \&
  {Bode}}]{SehgalTHB07}
{Sehgal}, N., {Trac}, H., {Huffenberger}, K., \& {Bode}, P. 2007, \apj, 664,
  149

\bibitem[{{Sharon} {et~al.}(2007){Sharon}, {Gal-Yam}, {Maoz}, {Filippenko}, \&
  {Guhathakurta}}]{SharonGMFG07}
{Sharon}, K., {Gal-Yam}, A., {Maoz}, D., {Filippenko}, A.~V., \&
  {Guhathakurta}, P. 2007, \apj, 660, 1165

\bibitem[{{Shimizu} {et~al.}(2004){Shimizu}, {Kitayama}, {Sasaki}, \&
  {Suto}}]{ShimizuKSS04}
{Shimizu}, M., {Kitayama}, T., {Sasaki}, S., \& {Suto}, Y. 2004, \pasj, 56, 1

\bibitem[{{Short} \& {Thomas}(2008)}]{ShortT08z}
{Short}, C.~J. \& {Thomas}, P.~A. 2008, ArXiv e-prints

\bibitem[{{Sijacki} {et~al.}(2007){Sijacki}, {Springel}, {di Matteo}, \&
  {Hernquist}}]{SijackiSMH07}
{Sijacki}, D., {Springel}, V., {di Matteo}, T., \& {Hernquist}, L. 2007,
  \mnras, 380, 877

\bibitem[{{Solanes} {et~al.}(2005){Solanes}, {Manrique}, {Gonz{\'a}lez-Casado},
  \& {Salvador-Sol{\'e}}}]{SolanesMGS05}
{Solanes}, J.~M., {Manrique}, A., {Gonz{\'a}lez-Casado}, G., \&
  {Salvador-Sol{\'e}}, E. 2005, \apj, 628, 45

\bibitem[{{Stanek} {et~al.}(2009){Stanek}, {Rudd}, \& {Evrard}}]{StanekRE09z}
{Stanek}, R., {Rudd}, D., \& {Evrard}, A.~E. 2009, \mnras, L172+

\bibitem[{{Staniszewski} {et~al.}(2008){Staniszewski}, {Ade}, {Aird}, {Benson},
  {Bleem}, {Carlstrom}, {Chang}, {Cho}, {Crawford}, {Crites}, {de Haan},
  {Dobbs}, {Halverson}, {Holder}, {Holzapfel}, {Hrubes}, {Joy}, {Keisler},
  {Lanting}, {Lee}, {Leitch}, {Loehr}, {Lueker}, {McMahon}, {Mehl}, {Meyer},
  {Mohr}, {Montroy}, {Ngeow}, {Padin}, {Plagge}, {Pryke}, {Reichardt}, {Ruhl},
  {Schaffer}, {Shaw}, {Shirokoff}, {Spieler}, {Stalder}, {Stark},
  {Vanderlinde}, {Vieira}, {Zahn}, \& {Zenteno}}]{Staniszewskiea08z}
{Staniszewski}, Z. et al. 2008, ArXiv e-prints

\bibitem[{{Strolger} {et~al.}(2004){Strolger}, {Dahlen}, {Livio}, {Panagia},
  {Challis}, {Tonry}, {Filippenko}, {Chornock}, {Ferguson}, {Koekemoer},
  {Mobasher}, {Dickinson}, {Giavalisco}, {Casertano}, {Hook}, {Blondin},
  {Leibundgut}, {Nonino}, {Rosati}, {Spinrad}, {Steidel}, {Stern}, {Garnavich},
  {Matheson}, {Grogin}, {Hornschemeier}, {Kretchmer}, {Laidler}, {Lee},
  {Lucas}, {de Mello}, {Moustakas}, {Ravindranath}, {Richardson}, \&
  {Taylor}}]{Strolger+04}
{Strolger}, L.-G. et al.  2004, \apj, 613, 200

\bibitem[{{Sullivan} {et~al.}(2006){Sullivan}, {Le Borgne}, {Pritchet},
  {Hodsman}, {Neill}, {Howell}, {Carlberg}, {Astier}, {Aubourg}, {Balam},
  {Basa}, {Conley}, {Fabbro}, {Fouchez}, {Guy}, {Hook}, {Pain},
  {Palanque-Delabrouille}, {Perrett}, {Regnault}, {Rich}, {Taillet}, {Baumont},
  {Bronder}, {Ellis}, {Filiol}, {Lusset}, {Perlmutter}, {Ripoche}, \&
  {Tao}}]{Sullivan+06}
{Sullivan}, M.  et al. 2006, \apj, 648, 868

\bibitem[{{Sun} {et~al.}(2009){Sun}, {Voit}, {Donahue}, {Jones}, {Forman}, \&
  {Vikhlinin}}]{SunVDJFV09}
{Sun}, M., {Voit}, G.~M., {Donahue}, M., {Jones}, C., {Forman}, W., \&
  {Vikhlinin}, A. 2009, \apj, 693, 1142

\bibitem[{{Suto} {et~al.}(1998){Suto}, {Sasaki}, \& {Makino}}]{SutoSM98}
{Suto}, Y., {Sasaki}, S., \& {Makino}, N. 1998, \apj, 509, 544

\bibitem[{{Tozzi} \& {Norman}(2001)}]{TozziNorman01}
{Tozzi}, P. \& {Norman}, C. 2001, \apj, 546, 63

\bibitem[{{Umetsu} {et~al.}(2008){Umetsu}, {Birkinshaw}, {Liu}, {Proty Wu},
  {Medezinski}, {Broadhurst}, {Lemze}, {Zitrin}, {Ho}, {Locutus Huang}, {Koch},
  {Liao}, {Lin}, {Molnar}, {Nishioka}, {Wang}, {Altamirano}, {Chang}, {Chang},
  {Chang}, {Chen}, {Han}, {Huang}, {Hwang}, {Jiang}, {Kesteven}, {Kubo}, {Li},
  {Martin-Cocher}, {Oshiro}, {Raffin}, {Wei}, \& {Wilson}}]{Umetsuea09z}
{Umetsu}, K.  2008, ArXiv e-prints

\bibitem[{{Vikhlinin}(2006)}]{Vikhlinin06}
{Vikhlinin}, A. 2006, \apj, 640, 710

\bibitem[{{Vikhlinin} {et~al.}(2009){Vikhlinin}, {Burenin}, {Ebeling},
  {Forman}, {Hornstrup}, {Jones}, {Kravtsov}, {Murray}, {Nagai}, {Quintana}, \&
  {Voevodkin}}]{VikhlininBEFHJKMNQV09}
{Vikhlinin}, A., {Burenin}, R.~A., {Ebeling}, H., {Forman}, W.~R., {Hornstrup},
  A., {Jones}, C., {Kravtsov}, A.~V., {Murray}, S.~S., {Nagai}, D., {Quintana},
  H., \& {Voevodkin}, A. 2009, \apj, 692, 1033

\bibitem[{{Vikhlinin} {et~al.}(2006){Vikhlinin}, {Kravtsov}, {Forman}, {Jones},
  {Markevitch}, {Murray}, \& {Van Speybroeck}}]{VikhlininKFJMMVS06}
{Vikhlinin}, A., {Kravtsov}, A., {Forman}, W., {Jones}, C., {Markevitch}, M.,
  {Murray}, S.~S., \& {Van Speybroeck}, L. 2006, \apj, 640, 691

\bibitem[{{Voit} \& {Bryan}(2001)}]{VoitBryan01}
{Voit}, G.~M. \& {Bryan}, G.~L. 2001, \apjl, 551, L139

\bibitem[{{Voit} {et~al.}(2002){Voit}, {Bryan}, {Balogh}, \&
  {Bower}}]{VoitBBB02}
{Voit}, G.~M., {Bryan}, G.~L., {Balogh}, M.~L., \& {Bower}, R.~G. 2002, \apj,
  576, 601

\bibitem[{{Wu} {et~al.}(2000){Wu}, {Fabian}, \& {Nulsen}}]{WuFN00}
{Wu}, K.~K.~S., {Fabian}, A.~C., \& {Nulsen}, P.~E.~J. 2000, \mnras, 318, 889

\bibitem[{{Yee} {et~al.}(2007){Yee}, {Gladders}, {Gilbank}, {Majumdar},
  {Hoekstra}, \& {Ellingson}}]{YeeGGMHE07}
{Yee}, H.~K.~C., {Gladders}, M.~D., {Gilbank}, D.~G., {Majumdar}, S.,
  {Hoekstra}, H., \& {Ellingson}, E. 2007, in Astronomical Society of the
  Pacific Conference Series, Vol. 379, Cosmic Frontiers, ed. N.~{Metcalfe} \&
  T.~{Shanks}, 103-

\end{thebibliography}


%
%
\begin{deluxetable}{lllllll}
\tabletypesize{\scriptsize}
\tablecaption{Decay times and yields\label{tab:starparams}}
\tablewidth{0pt}
\tablehead{
\colhead{mass}                                  &
\colhead{mean}                                  &
\colhead{remnant}                               &
\colhead{remnant}                               &
\colhead{$\eta_i$}                              &
\colhead{$\nu_i$}                             &
\colhead{$\tau_i$}                                        \\  
\colhead{range}                                 &
\colhead{mass}                                  &
\colhead{type}                                  &
\colhead{mass}                                  &
\colhead{(fraction)}                           &
\colhead{$(M_\odot^{-1})$}             &
\colhead{(Gyr)}                                  } 
\startdata
50-100& 68.4 & BH      & 2.0 & 1   & $5.11\times 10^{-4}$ & 0.01  \\
8-50  & 16.0 & SNCC+NS & 1.4 & 1   & $9.14\times 10^{-3}$ & 0.01   \\
3-8   & 4.58 &prompt IA& 0   &0.07 & $1.93\times 10^{-3}$ & 0.05   \\
3-8   & 4.58 & tardy IA& 0   &0.07 & $1.93\times 10^{-3}$ & 3.01   \\
3-8   & 4.58 & WD      &0.55 &0.92 & 0.0253               & 0.5    \\
1.1-3 & 1.69 & WD      &0.55 & 1   & 0.1078               & 6.0    
\enddata
\tablecomments{Masses in $M_\odot$.}
\end{deluxetable}
%
%


\epsscale{0.93}

\epsscale{0.45}

\end{document}